\documentclass[10pt,prb,aps,twocolumn,floatfix,nofootinbib]{revtex4-2}
\usepackage{graphicx}
\usepackage{amssymb}
\usepackage{amsmath}

\begin{document}

\author{Anders W. Sandvik}
\affiliation{Department of Physics, Boston University, 590 Commonwealth Avenue, Boston, Massachusetts 02215}
\affiliation{School of Physical and Mathematical Sciences, Nanyang Technological University, Singapore}

\title{High-precision ground state parameters \\ of the two-dimensional spin-1/2 Heisenberg model on the square lattice}

\begin{abstract}
Motivated by the need for high-precision benchmark results for basic quantum many-body models, several ground state properties of
the square-lattice $S=1/2$ Heisenberg antiferromagnet are computed (the energy, order parameter, spin stiffness, spinwave velocity,
long-wavelength susceptibility, and staggered susceptibility) using extensive quantum Monte Carlo simulations with the stochastic
series expansion method. Moderately sized lattices are studied at temperatures $T$ sufficiently low to realize the $T \to 0$ limit,
with significantly reduced statistical errors relative to previous works. Results for periodic $L\times L$ lattices with $L \in [6,96]$
are tabulated versus $L$ and extrapolations to infinite system size are carried out. The relative statistical errors of the finite-size
energies are below $10^{-7}$. The extrapolated ground state energy density is $e_0=-0.669441857(7)$, where the number
within parentheses is the statistical error (one standard deviation) of the preceding digit. This value represents an improvement
in precision of three orders of magnitude over the previously best result. The leading and subleading finite-size corrections to
$e_0$ are in full quantitative agreement with predictions from chiral perturbation theory, thus further supporting the soundness of both
the extrapolations and the theory. The extrapolated sublattice magnetization is $m_s=0.307447(2)$, which agrees well with previous estimates
but with a much smaller statistical error. The coefficient of the linear in $L^{-1}$ correction to $m^2_s$ agrees with the value from
chiral perturbation theory and the presence of a factor $\ln^\gamma(L)$ in the second-order correction is also confirmed, with the previously
not known value of the exponent being $\gamma = 0.82(4)$. The finite-size corrections to the staggered susceptibility point to logarithmic
corrections also in this quantity. To facilitate benchmarking of methods for which periodic boundary conditions are challenging,
results for systems with open and cylindrical boundaries are also listed and their spatially inhomogeneous order parameters are analyzed.
\end{abstract}

\date{\today}

\maketitle

\section{Introduction}
\label{sec:intro}

In light of ongoing efforts to develop better computational methods for strongly interacting quantum many-body systems, accurate benchmark
results for key models are called for. While quantum Monte Carlo (QMC) simulations are limited to models for which the sign problem
\cite{loh90,henelius00,troyer05} can be avoided, these are the only numerically (statistically) exact methods that can reach arbitrarily low
temperatures $T$ for large systems in spatial dimensionality higher than one. In particular, with the stochastic series expansion (SSE) QMC
method \cite{sandvik99,syljuasen02,sandvik10b} with loop updates \cite{evertz93} and related ground-state projection techniques \cite{sandvik10a},
two-dimensional (2D) $S=1/2$ quantum spin models have been studied in the $T \to 0$ limit on lattices with linear size $L$ up to hundreds of
lattice constants \cite{ma18,sandvik20} or even larger in some cases \cite{takahashi24}. These system sizes far exceed the current limitations
of more versatile ground state techniques, such as the density matrix renormalization (DMRG) method \cite{white93,schollwoeck11,stouden13}, a variety of methods
based on tensor products (TPs) \cite{murg09,liu18,orus19} (which also can be formulated for infinite lattices, where there are still other convergence
issues \cite{corboz25}), and, more recently, neural networks (NNs) \cite{carleo17,lange24}. These emerging approaches essentially produce sophisticated
variational states, where the number of optimized parameters in principle can be taken sufficiently large to converge to the true ground state.
In practice, there are still formidable challenges with the scaling of the computational effort (both memory and time) with all these methods.
The ability to obtain correct results can then be fairly judged only by comparing with exact solutions or QMC simulations of sign-free models,
often with the 2D Heisenberg model as a first non-trivial test beyond DMRG accessible one-dimensional systems
\cite{wu23,chen24,rende24,bortone24,wang24,kull24,moss25}.

The primary motivation for the present work is to provide improved QMC benchmark results for the ground state of the standard 2D $S=1/2$ Heisenberg
antiferromagnet, described by the Hamiltonian
\begin{equation}
H = \sum_{\langle ij\rangle}{\bf S}_i \cdot {\bf S}_j,
\end{equation}
where the coupling has been set to unity and $\langle ij\rangle$ represents nearest-neighbor spins on the square lattice with $L \times L$ sites and even $L$.
Most of the results will be for periodic boundary conditions but some quantities obtained with open and cylindrical boundaries will also be listed, to facilitate
benchmarking of methods for which calculations with periodic boundaries are challenging. In addition to lists of observables versus the lattice size,
$L \to \infty$ extrapolated values will also be presented. Detailed comparisons will be made with finite-size corrections previously predicted using
chiral perturbation theory \cite{hasen93,niedermayer11} in the case of periodic $L \times L$ lattices. For the other boundary conditions, the spatially
inhomogeneous order parameter will be analyzed.

The Heisenberg model has been studied in many past works with a variety of methods and the nature of the N\'eel ordered ground state is now well established
\cite{anderson59,halperin69,oitmaa78a,oitmaa78b,huse88,reger88,chn89,liang90,makivic91,manousakis91,runge92a,runge92b,wiese94,beard96,sandvik97,jiang11,sen15}.
Over time, the statistical precision achieved in QMC simulations has increased to a level necessary for detailed tests \cite{wiese94,beard96,sandvik97,sandvik10a,jiang11,sen15}
of field-theoretical descriptions \cite{chn89,neuberger89,fisher89,hasen93,niedermayer11} of the N\'eel state. However, TN and NN methods are now in some
cases reaching lower estimated variational energy errors for periodic lattices (which are difficult to handle with traditional DMRG methods) than the
statistical errors of the best published QMC benchmark results \cite{chen24}. While these methods ultimately target models with interactions beyond the
sign-free bipartite Heisenberg exchange, in particular geometrically frustrated systems, it will still be useful to have improved QMC data for the
plain 2D Heisenberg model.

A second motivation for improving on previous results is the continued interest in testing universal scaling behaviors related
to the breaking of the SO(3) spin-rotation symmetry, e.g., as predicted in great detail by chiral perturbation theory. The finite-size values of the ground
state parameters studied here should be polynomials in the inverse system size \cite{fisher89,neuberger89,hasen93,niedermayer11} with coefficients that
themselves depend on the infinite-size parameters. In the case of the sublattice magnetization, a logarithmic (log) correction has also been pointed out
\cite{neuberger89}. The predicted coefficients of both leading and subleading corrections are successfully tested here to significantly higher precision
than previously, including the aforementioned log correction for which a previously not known exponent is determined.

The QMC results reported here were obtained with the SSE method at sufficiently low temperatures to realize the $T \to 0$ limit. The goal is not to reach
the largest possible system sizes but to consider small to moderate lattices and reduce the statistical errors significantly with respect to those in
Refs.~\cite{sandvik97} and \cite{sandvik10a}, which have become standard references for the ground state energy and spin correlations of the 2D Heisenberg model.
Other quantities computed here include the spin stiffness, the long-wavelength susceptibility, the spinwave velocity, and the staggered susceptibility. These
quantities are known to converge slower than the energy in TP and NN methods and are in some cases difficult to calculate. They should nevertheless also be subjected
to rigorous benchmarking alongside variational energy estimates.

The largest system size included in the SSE calculations in Ref.~\cite{sandvik97} was $L=16$, which is here extended up to $L=96$. The error reduction
in some observables, in particular the energy, is by orders of magnitude for both the finite-size data and $L \to \infty$ extrapolated values. The
extrapolated energy when constraining fits by chiral perturbation theory input (after testing the consistency of those predictions) is $e_0=-0.669441857(7)$
while the previously best estimate was $e_0=-0.669437(5)$. In the case of the order parameter, lattice sizes up to $L=256$ had been previously studied using a ground
state projector method in the valence-bond basis \cite{sandvik10a}, with the extrapolated result $m_s=0.30743(1)$. An identical result was obtained by fitting
both finite-size and finite-temperature QMC results to forms predicted by chiral perturbation theory \cite{jiang11}. Here the new results for $L\le 96$ are
combined with the previous $m^2_s(L)$ results for $L=128,192$ and $256$ \cite{sandvik10a} and extrapolated using a fit including the predicted log correction
to the $L^{-2}$ term (the leading correction being linear in $L^{-1}$). This form without any higher-order corrections fits the data for $L \ge 12$ extremely
well, perhaps suggesting that there is no $L^{-3}$ correction. All data for $L \ge 6$ can be fitted with an added $L^{-5}$ correction, which leads to the best
extrapolated sublattice magnetization, $m_s=0.307447(2)$; by far the most precise estimate to date.

The rest of the paper is organized as follows: In Sec.~\ref{sec:finite}, after defining the relevant physical observables, the ground-state convergence
versus the inverse temperature is discussed and tested. Converged results for different system sizes are tabulated. The observables are extrapolated to
the thermodynamic limit in Sec.~\ref{sec:infinite} and the validity of the forms of the finite-size corrections predicted by chiral perturbation theory
is tested. Results for lattices with open and cylindrical boundary conditions are listed in Sec.~\ref{sec:bc} and the edge distortion of the order
parameter is also analyzed. Sec.~\ref{sec:discuss} concludes with a brief summary and discussion.

\section{Finite-Size results}
\label{sec:finite}

The ground state properties considered here were computed within the efficient SSE operator-loop approach \cite{sandvik99,sandvik10b}, where the
partition function is expressed as
\begin{equation} \label{zsse}
Z = {\rm Tr}\{ {\rm e}^{-\beta H} \} = \sum_{n=0}^\infty \frac{(-\beta)^n}{n!} {\rm Tr}\{H^n \},  
\end{equation}
at inverse temperature $\beta$. Details of the method and estimators of various observables are described in detail in the existing literature
\cite{sandvik97,sandvik10b}, and neither algorithms nor technical implementations will be significantly discussed here. The SSE method itself is formally
exact and the results are only affected by standard statistical sampling errors, characterized in the results to be presented here as one standard deviation
of the mean. To obtain the ground state parameters of interest, the limit $T \to 0$ also must be carefully taken. Definitions of the physical observables
will be given first in Sec.~\ref{sec:defs}, followed by $T \to 0$ convergence checks in Sec.~\ref{sec:conv} and finite-lattice results
in Sec.~\ref{sec:resfinite}.

\subsection{Definitions}
\label{sec:defs}

The internal energy density [with a constant subtracted from $H$ in Eq.~(\ref{zsse}) to make the expansion positive definite]
is obtained in SSE simulations as the mean of the expansion power $n$ in Eq.~(\ref{zsse}),
\begin{equation} \label{esse}
e = \frac{\langle H\rangle}{N} = \frac{\langle n\rangle}{\beta N},
\end{equation}
where $N=L^2$ is the system volume. Not being explicitly dependent on the evolved spin configurations, this is manifestly a spin-rotationally averaged
estimator \cite{sandvik97}. The width of the distribution of $n$ scales as $\langle n \rangle^{1/2}$ for $T \to 0$, which together with the above SSE form of
$e$ implies that even a single instance of $n$ gives an energy estimate that is good to within an error of order only $(\beta N)^{-1/2}$. For $M$ independent
sampled SSE configurations, the statistical error is therefore of order  $(\beta NM)^{-1/2}$, i.e., for increasing $\beta$ and $N$ less sampling is required to
achieve a specific statistical precision. For desired precision (standard deviation of the mean) $\epsilon$, the number of independent measurements must
be of order $(\epsilon^2\beta N)^{-1}$. Assuming no $N$ and $\beta$ dependence of the autocorrelation time of $n$, the computational effort for generating
an independent sample of $n$ scales as $\beta N$ (in terms of the number of operations performed, disregarding computer technical aspects such as worse cache
memory performance for increasing system size), thus, the effort to achieve a desired precision of $e$ would be independent of $\beta$ and $N$. The SSE
autocorrelation time of $n$ is indeed short and shows almost no dependence on $\beta$ and $N$, and the above scaling behavior of $e$
seems to be satisfied in practice. The other size-normalized quantities listed below typically exhibit increasing autocorrelation
times with $N$, and in some cases with $\beta$, which is reflected in increasing relative statistical errors with $L$ in the results to be
presented below. The total lengths of the SSE simulation runs were adapted to achieve a relative error in $e(L)$ of roughly $10^{-7}$
for each system size considered.

The following is just a concise list of the other quantities considered, with only minimal comments on how they are estimated within the
SSE method. For further technical details, see Refs.~\cite{sandvik97} and \cite{sandvik10b}.

The order parameter density is the staggered magnetization, which can be taken as a diagonal operator in the SSE simulations,
\begin{equation} \label{mzsdef}
 m^z_s = \frac{1}{N}\sum_{i=1}^N (-1)^{x_i+y_i}S^z_i,
\end{equation}
where $(x_i,y_i)$ are the integer lattice coordinates. The corresponding extensive quantity lacking the $N^{-1}$ factor will also be referred
to below; $M^z_s=Nm_s^z$. The definition, \begin{equation} \label{ms2def}
 m^2_s \equiv 3\langle (m^z_s)^2\rangle,
\end{equation}
includes a factor $3$ to account for spin-rotational averaging, so that the full symmetry-broken value is obtained in the thermodynamic limit.
The expectation value was computed with an efficient loop estimator \cite{evertz93,wiese94,sandvik10b}, which, however, was not fully averaged
over imaginary time in order to reduce the computational effort of these measurements---given that one of the main aims here is the obtain
high-precision results for the energy, the estimator of which demands no computational effort.

The corresponding staggered susceptibility is given by the Kubo integral
\begin{equation}\label{ssusc}
\chi_s = \frac{1}{N}\int_0^\beta d\tau \langle M^z_s(\tau)M^z_s(0)\rangle,
\end{equation}  
where $M^z_s(\tau)$ is the imaginary-time evolved total sublattice magnetization. This space-time integrated correlator
can also in principle can be computed with a loop estimator \cite{beard96} but here a different approach was taken, where an exact
expression for the integral is evaluated exactly for each SSE configuration \cite{sandvik10b,sandvik91} but is not expressed using loop sizes. 
Though the loop estimator should in principle be more efficient, the results obtained here are still of high statistical quality.

The staggered magnetization corresponds to wavenumber ${\bf q}=(\pi,\pi)$ and Eq.~(\ref{ssusc}) generalizes to arbitrary ${\bf q}$;
\begin{equation}\label{qsusc}
\chi({\bf q}) = \frac{1}{N}\int_0^\beta d\tau \langle M^z(-{\bf q},\tau)M^z({\bf q},0)\rangle,
\end{equation}  
where $M^z({\bf q},\tau)$ is the Fourier transform of the spins at fixed imaginary time (here not normalized by $N$), i.e.,
\begin{equation}
 M^z({\bf q},\tau) = \sum_{i=1}^N {\rm e}^{-{\bf r}_i \cdot {\bf q}} {\rm e}^{\tau H}S^z_i {\rm e}^{-\tau H},
\end{equation}
with ${\bf r}_i = (x_i,y_i)$. For $q=0$, $M^z({\bf q},\tau)$ is the conserved uniform magnetization and the Kubo integral then reduces
to the classical form of the uniform susceptibility
\begin{equation} \label{ususc}
\chi_u = \frac{1}{NT}\left ( \sum_{i=1}^N S^z_i\right )^2,
\end{equation}
which vanishes identically in the singlet ground state. The long-wavelength susceptibility is considered here by using ${\bf q}_1=(2\pi/L,0)$ or
${\bf q}_1=(0,2\pi/L)$ in Eq.~(\ref{qsusc}). In the gapless symmetry broken state, the transverse component $\chi_\perp$ of this susceptibility is a key
ground state parameter, which can be obtained in the large-$L$ limit from $\chi(q_1)$ by including a rotational factor $3/2$ (reflecting the fact that the
longitudinal component should vanish as $T \to 0$ on account of the conserved magnetization and gapped longitudinal fluctuations):
\begin{equation}\label{psusc}
\chi_\perp = \frac{3}{2}\chi(q_1).
\end{equation}  

The spin stiffness for systems defined on periodic lattices is computed using the total spin current (winding number) fluctuations \cite{pollock87,sandvik97} as
\begin{equation} \label{rhosdef}
\rho_s = \frac{3}{4\beta N}\langle W_x^2 + W_y^2\rangle,
\end{equation}
where $W_a$, $a\in \{x,y\}$, is the current along the $a$ lattice direction integrated over imaginary time (which is done exactly
with the SSE estimator),
\begin{equation}
W_a = \int_0^\beta d\tau J_a(\tau),
\end{equation}
which is an integer multiple (the winding number) of the system length $L$ in a periodic system. Eq.~(\ref{rhosdef}) also contains a factor
$3/2$ to account for the difference between rotationally invariant finite systems and the symmetry-broken state in the thermodynamic limit.

The spinwave velocity can be obtained from the stiffness and the transverse susceptibility using a formula originating from the hydrodynamic
description \cite{halperin69} of the long-wavelength fluctuations of the N\'eel state,
\begin{equation}\label{chydro}
c=\sqrt{\frac{\rho_s}{\chi_\perp}},
\end{equation}
which should be exact when the $L \to \infty$ limit is taken of $\rho_s$ and $\chi(q_1)$. Finite-size results $c(L)$ based on the same formula
will also be presented here.

\begin{figure}[t]
\includegraphics[width=80mm]{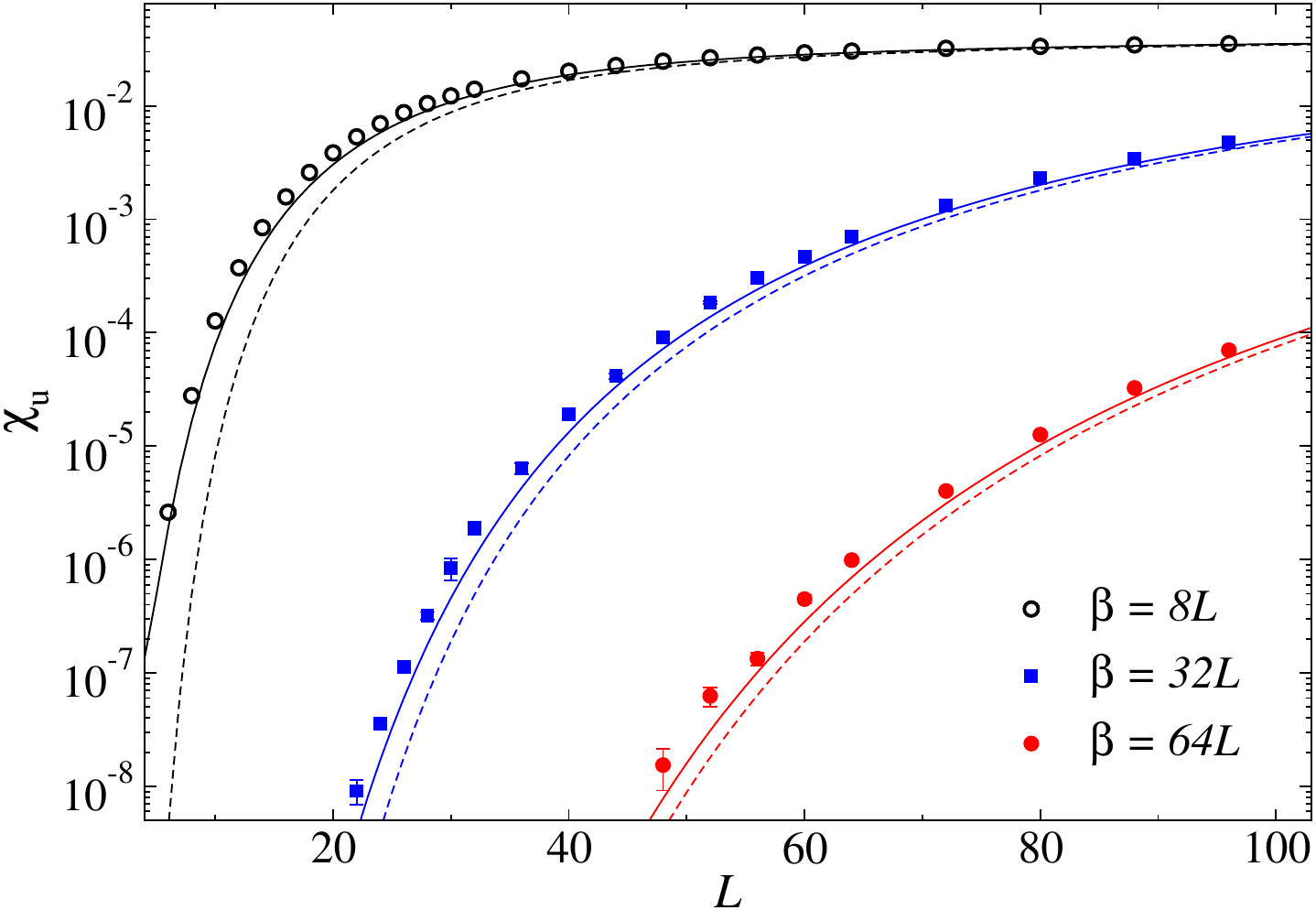}
\caption{Uniform susceptibility versus the system size for three values of the aspect ratio $\beta/L$. The dashed lines show predictions
based on a quantum rotor tower with the infinite-size value of the $T=0$ susceptibility, $\chi_u(\infty)=0.04379$, while the solid lines show
results with $\chi_u(L)$ given by a fourth-order polynomial fitted to the finite-size data in Table \ref{ltable}.}
\label{xu_beta}
\end{figure}

\subsection{Finite-temperature effects and $T \to 0$ convergence}
\label{sec:conv}

In order to judge remaining effects of finite temperature in the SSE simulations, the tower of Anderson quantum rotor excitations \cite{anderson59} can be
analyzed. For the 2D system, these excitations with total spin $S > 0$ (the ground state having $S=0$) should have the exact asymptotic form 
\begin{equation} \label{erotor}
E(S)=\frac{S(S+1)}{2\chi_\perp L^2}
\end{equation}
for $S$ up to order $L^{1/2}$ and large $L$ (and with $L^2 \to L^d$ for a $d$-dimensional system). These rotor states fall below the lowest spinwave excitation
(the energy of which is of order $L^{-1}$) and become degenerate with the ground state as $L \to \infty$, thus allowing for the spontaneous breaking of the
SO(3) symmetry of the  N\'eel order parameter. A pedagogical derivation of the numerator in Eq.~(\ref{erotor}) can be found in Ref.~\cite{sandvik10b}.

The $L^{-2}$ scaling of the rotor excitations immediately implies that full convergence to the ground state requires the inverse temperature to
grow with $L$ as $\beta \propto L^2$. In practice, convergence can be tested by monitoring computed expectation values versus $\beta$. It is
then useful to pay particular attention to quantities that are expected to converge slowly, e.g., the staggered susceptibility Eq.~(\ref{ssusc}),
where the integral over $\tau$ is of a very slowly decaying correlator and $\beta$ has to be large not only to converge equal-time properties but
also to accommodate sufficiently large $\tau$ for the correlations to vanish exponentially, ultimately as ${\rm e}^{-E(S)\tau}$ with $S=1$ in Eq.~(\ref{erotor}).

Another useful quantity is the uniform susceptibility Eq.~(\ref{ususc}), which vanishes identically in the singlet ground state but is non-zero at
any $T>0$, ultimately on account of the $S>0$ rotor states at very low $T$. Fig.~\ref{xu_beta} shows the susceptibility versus the system size when
$\beta$ is scaled only linearly with $L$ as $\beta=aL$, but with rather large factors $a=8$, $32$, and $64$. The results are compared with the prediction
from the rotor spectrum Eq.~(\ref{erotor});
\begin{equation}
\chi_u(T)= \frac{1}{Z} \sum_{S=1}^L {\rm e}^{-E(S)/T}\sum_{m=-S}^S m^2,
\end{equation}
where the partition function is
\begin{equation}\label{partfun}
Z = 1+\sum_{S} (2S+1){\rm e}^{-E(S)/T}.
\end{equation}
Here the upper bound for $S$ should be of order $L^{1/2}$, reflecting the scale at which the rotor energy is comparable to the lowest spinwave energy.
However, at the low temperatures of interest here the expectation value is completely dominated by $S \ll L^{1/2}$ and the exact value of the cutoff is not
important (it is taken all the way to $S=L$ in the calculations here). The value of the perpendicular susceptibility in Eq.~(\ref{erotor}) is here taken
from the infinite-size polynomial extrapolation (discussed in Sec.~\ref{sec:infinite}) of the data in Table \ref{ltable}, which results in the dashed curves
in Fig.~\ref{xu_beta}. Though there are clearly deviations, the predictions approach the SSE results as $L$ increases. To account for some of the finite-size
corrections, $\chi_\perp$ can alternatively be treated as a size dependent quantity $\chi_\perp(L)$, which can be evaluated for arbitrary $L$ using the
same fitted polynomial as used for the $L \to \infty$ extrapolation. As shown with the solid curves in Fig.~\ref{xu_beta}, the deviations when using 
$\chi_\perp(L)$ are even smaller and leave little doubt that the rotor tower provides a good (asymptotically exact) description for temperatures
below the lowest spinwave excitations. The deviations in Fig.~\ref{xu_beta} from the predicted curves can be explained by finite-size corrections
to the rotor spectrum. These corrections have been studied in previous works \cite{lavalle98,syljuasen02,sandvik10b,nieder18}.

\begin{figure}[t]
\includegraphics[width=80mm]{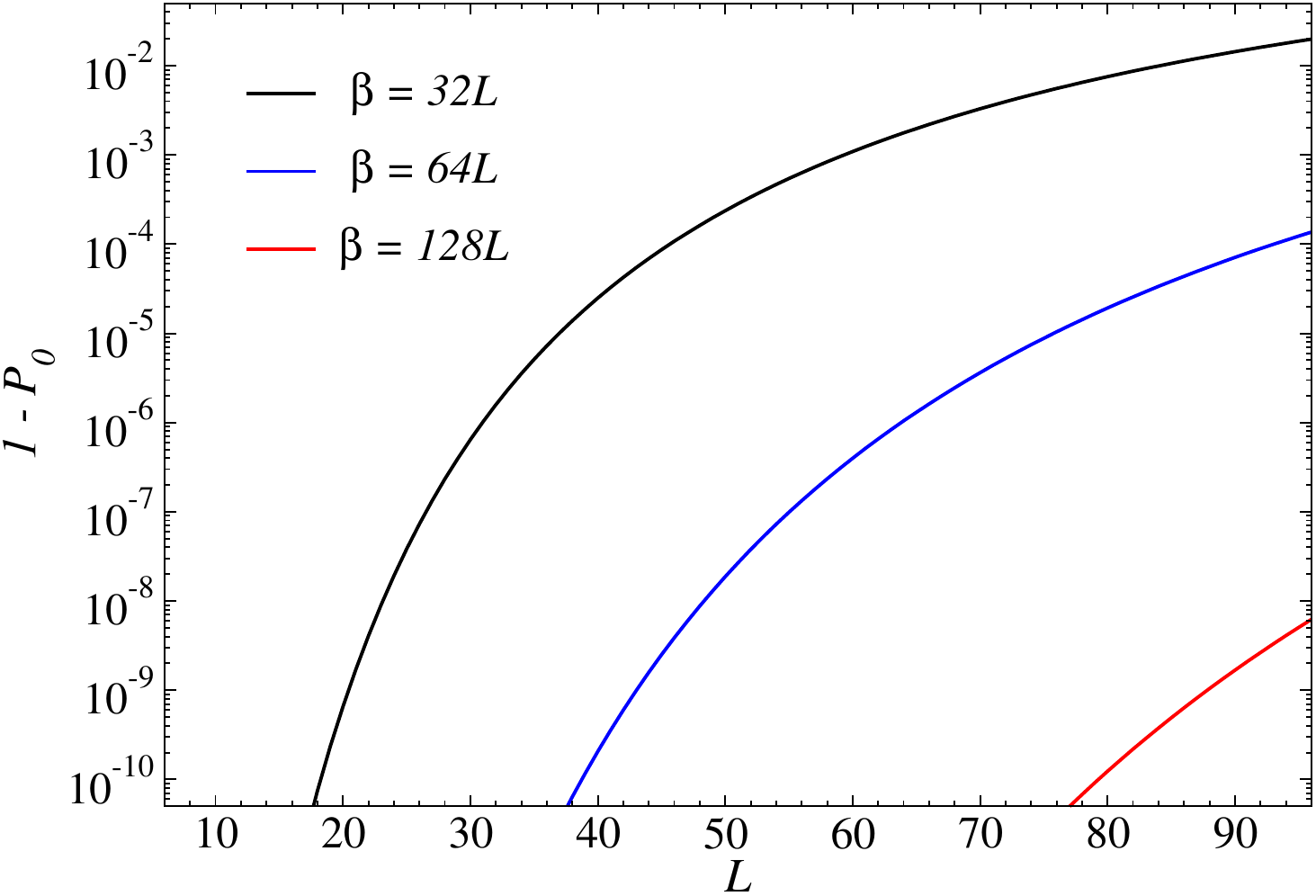}
\caption{Probability vs the lattice size of the system being thermally excited ($P_0$ being the ground-state probability) under the valid assumption
that only the rotor states contribute. The results for $\beta=32$ (black), $64$ (blue) and $128$ (red) were computed using the interpolated $\chi_\perp(L)$
function in the rotor spectrum Eq.~(\ref{erotor}), the same as used for the solid curves in Fig.~\ref{xu_beta}. The range of $L \in [6,96]$ corresponds
to that for which SSE results are presented here.}
\label{p0_beta}
\end{figure}

From the partition function Eq.~(\ref{partfun}) we can also compute the probability $P_0$ of the system being in the ground state. Given
that the weight of the ground state is $1$ in Eq.~(\ref{partfun}), we simply have $P_0=Z^{-1}$. The corresponding probability $1-P_0$ of the
system being excited into the rotor tower is graphed versus the system size at $\beta/L=32, 64$, and $128$ in Fig.~\ref{p0_beta}. Convergence
of computed observables were confirmed explicitly by comparing results at different $\beta$ values, but the results in Fig.~\ref{p0_beta} are
also useful for judging whether results of the two different calculations can be averaged and still effectively maintain the $T=0$ limit.
Most physical observables change very little, in a relative sense, between the ground state and a low-lying excited state, e.g., the energy density
of a rotor state is only of order $N^{-2}$ above the ground state. Therefore, the relative finite-temperature corrections in such observables will
be orders of magnitude smaller than the excitation probabilities in Fig.~\ref{p0_beta}.

It should be noted here that simulations with $\beta$ larger than what is strictly needed to obtain $T \to 0$ results is not a waste of computer
resources, because the amount of statistical data contained in the SSE configurations scales linearly with $\beta$, i.e., in the same way as the nominal
computational effort of the simulations---the example of the energy was already discussed explicitly at the beginning of Sec.~\ref{sec:finite}.
An excessively large $\beta$ can still be detrimental in practice because of memory limitations and poor cache performance. The calculations
reported here were mostly limited to $\beta/L \le 64$ and system sizes $L \le 96$, with some limited test results generated also at $\beta/L=128$
for $L=80,88$, and $96$.

In this work the $T \to 0$ limit is effectively taken first for each system size $L$, with results presented below in Sec.~\ref{sec:resfinite}.
Such ground state data will be useful for benchmarking other methods, as discussed in Sec.~\ref{sec:intro}. The $L \to \infty$ limit of the various
observables will be considered in Sec.~\ref{sec:infinite} by fitting the finite-$L$ data sets to polynomials in $L^{-1}$---the predicted form of the
finite-size corrections (in one case with a log correction) \cite{neuberger89,hasen93,niedermayer11} that will be tested as one of the aims
of this work.

It is worth pointing out that the limits $L \to \infty$ and $T \to 0$ can be taken in either order for most quantities that do not diverge when
$T \to 0$. A prominent exception is the uniform susceptibility, Eq.~(\ref{ususc}), on account of the vanishing conserved magnetization as $T \to 0$
for any $L$. However, the uniform magnetization can also be taken as the limit $q \to 0^+$, in practice with $q_1=2\pi/L$, of the general susceptibility
in Eq.~(\ref{qsusc}), which does not vanish when $T \to 0$ for any $L$. The limit can then safely be taken with $T \to 0$ data versus $L$.

\begin{figure}[t]
\includegraphics[width=80mm]{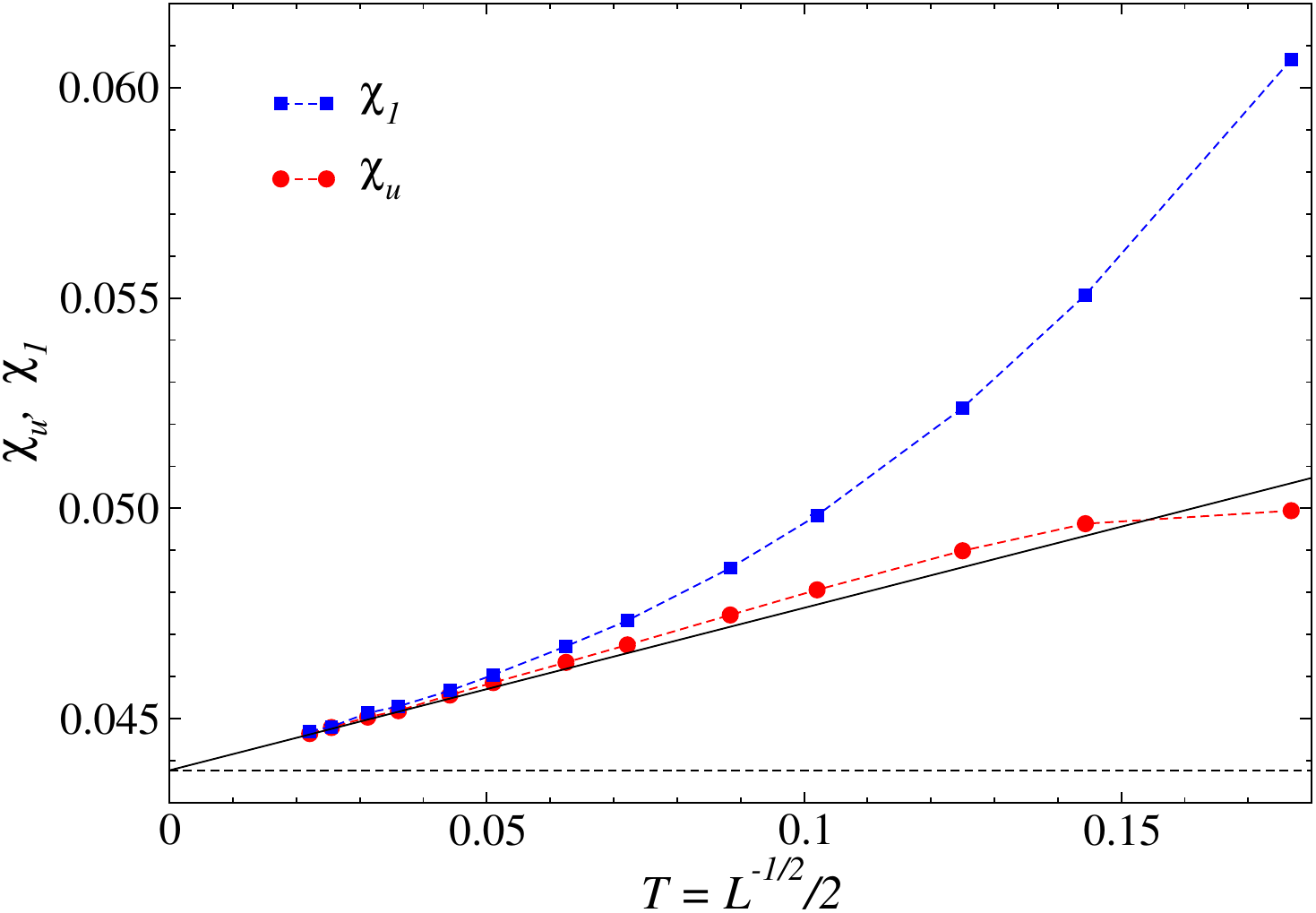}
\caption{Temperature dependent $q\to 0$ susceptibility computed in two different ways in simulations with $\beta=2L^{1/2}$; with $q=0$ in $\chi_u$ and
$q=2\pi/L$ in $\chi_1$. The dashed line is the $T=0$ value obtained by extrapolating the $\chi_\perp$ data in Table \ref{ltable} to the thermodynamic
limit and adjusted for the factor $3/2$ in Eq.~(\ref{qsusc}). The black solid line is a linear fit to the $\chi_u$ data. The other dashed lines are guides
to the eye.}
\label{xt}
\end{figure}

Though we will here not use alternative approaches to taking the limit $L \to \infty$ first and then $T \to 0$, it is interesting to note that the two
limits can also in principle be taken simultaneously as a function of a single parameter, by considering a size dependent temperature $T(L)$, adapted such
that a sufficient fraction of the quantum rotor states fall below this temperature. A natural choice in terms of the inverse temperature is
$\beta = aL^{1/2}$, with $a$ an essentially arbitrary constant (with value only relevant as far as it determines the rate of convergence to the $T\to 0$
limit). Fig.~\ref{xt} shows results versus the temperature for both $\chi_u$ and $\chi_1=\chi(q_1)$ obtained with $\beta=2L^{1/2}$. Here the equality
of the two susceptibilities in the limit $T \to 0$ can be observed clearly.

\begin{table*}[t]
\caption{Results for different system sizes computed at inverse temperature $\beta/L=64$ for $L > 30$ and averaged over $\beta/L=32$
and $\beta/L=64$ simulations for $L \le 30$. The digits within parentheses indicate the statistical error (one standard deviation rounded to
a single digit) of the preceding digit of the mean.}
\begin{tabular}{|r|l|l|l|l|l|l|}
\toprule    ~~$L$~ & ~~~~~~$e_0$ & ~~~~~~$\rho_s$ & ~~~~~~$\chi_\perp$ & ~~~~~~$c$ & ~~~~~~$m_s^2$ & ~~~~~~$\chi_s/L^2$ \cr
\hline
~ 6~&~ -0.67887215(2)~&~  0.2375897(1)~&~ 0.0936076(1)~&~ 1.593155(1)~&~ 0.2098368(2)~&~ 0.01324669(3)~
\cr \hline
~ 8~&~ -0.67349005(2)~&~  0.2205407(1)~&~ 0.0854012(1)~&~ 1.606987(1)~&~ 0.1778419(2)~&~ 0.01038140(3)~
\cr \hline
~10~&~ -0.67155266(2)~&~  0.2112378(1)~&~ 0.0806085(2)~&~ 1.618809(2)~&~ 0.1593710(2)~&~ 0.00883242(4)~
\cr \hline
~12~&~ -0.67068192(2)~&~  0.2054239(1)~&~ 0.0775450(2)~&~ 1.627604(2)~&~ 0.1474495(3)~&~ 0.00788157(4)~
\cr \hline
~14~&~ -0.67023225(2)~&~  0.2014610(2)~&~ 0.0754479(3)~&~ 1.634075(3)~&~ 0.1391544(4)~&~ 0.00724419(5)~
\cr \hline
~16~&~ -0.66997660(2)~&~  0.1985904(2)~&~ 0.0739352(3)~&~ 1.638904(4)~&~ 0.1330657(4)~&~ 0.00678935(6)~
\cr \hline
~18~&~ -0.66982043(2)~&~  0.1964164(3)~&~ 0.0728004(4)~&~ 1.642562(5)~&~ 0.1284152(5)~&~ 0.00644930(7)~
\cr \hline
~20~&~ -0.66971967(2)~&~  0.1947161(3)~&~ 0.0719222(5)~&~ 1.645393(5)~&~ 0.1247499(6)~&~ 0.00618556(7)~
\cr \hline
~22~&~ -0.66965176(2)~&~  0.1933480(4)~&~ 0.0712254(5)~&~ 1.647603(7)~&~ 0.1217905(7)~&~ 0.00597567(9)~
\cr \hline
~24~&~ -0.66960434(2)~&~  0.1922247(4)~&~ 0.0706606(6)~&~ 1.649361(7)~&~ 0.1193517(7)~&~ 0.00580444(9)~
\cr \hline
~26~&~ -0.66957017(3)~&~  0.1912848(6)~&~ 0.0701934(8)~&~ 1.65079(1)~&~  0.1173082(9)~&~ 0.0056624(1)~
\cr \hline
~28~&~ -0.66954496(3)~&~  0.1904871(7)~&~ 0.0698049(9)~&~ 1.65192(2)~&~  0.115573(1)~&~  0.0055424(1)~
\cr \hline
~30~&~ -0.66952595(3)~&~  0.1898036(8)~&~ 0.069472(1)~&~  1.65290(2)~&~  0.114080(1)~&~  0.0054402(2)~
\cr \hline
~32~&~ -0.66951133(3)~&~  0.1892071(9)~&~ 0.069188(1)~&~  1.65369(2)~&~  0.112781(1)~&~  0.0053517(2)~
\cr \hline
~36~&~ -0.66949091(3)~&~  0.188225(1)~&~  0.068725(2)~&~  1.65494(2)~&~  0.110638(2)~&~  0.0052065(2)~
\cr \hline
~40~&~ -0.66947776(3)~&~  0.187448(2)~&~  0.068368(2)~&~  1.65583(2)~&~  0.108941(2)~&~  0.0050922(2)~
\cr \hline
~44~&~ -0.66946889(3)~&~  0.186819(2)~&~  0.068083(2)~&~  1.65650(2)~&~  0.107570(2)~&~  0.0050002(3)~
\cr \hline
~48~&~ -0.66946274(3)~&~  0.186297(2)~&~  0.067855(2)~&~  1.65696(3)~&~  0.106426(3)~&~  0.0049242(3)~
\cr \hline
~52~&~ -0.66945833(3)~&~  0.185860(2)~&~  0.067661(2)~&~  1.65738(3)~&~  0.105474(3)~&~  0.0048611(3)~
\cr \hline
~56~&~ -0.66945502(3)~&~  0.185483(2)~&~  0.067501(3)~&~  1.65767(4)~&~  0.104660(3)~&~  0.0048066(3)~
\cr \hline
~60~&~ -0.66945260(3)~&~  0.185158(3)~&~  0.067367(3)~&~  1.65786(4)~&~  0.103955(3)~&~  0.0047602(4)~
\cr \hline
~64~&~ -0.66945072(3)~&~  0.184876(3)~&~  0.067245(3)~&~  1.65810(4)~&~  0.103342(4)~&~  0.0047199(4)~
\cr \hline
~72~&~ -0.66944815(3)~&~  0.184410(4)~&~  0.067052(4)~&~  1.65839(5)~&~  0.102330(4)~&~  0.0046532(5)~
\cr \hline
~80~&~ -0.66944645(3)~&~  0.184040(5)~&~  0.066905(4)~&~  1.65854(6)~&~  0.101527(5)~&~  0.0046004(5)~
\cr \hline
~88~&~ -0.66944530(3)~&~  0.183736(6)~&~  0.066774(5)~&~  1.65880(7)~&~  0.100864(5)~&~  0.0045559(6)~
\cr \hline
~96~&~ -0.66944454(3)~&~  0.183480(7)~&~  0.066674(6)~&~  1.65888(8)~&~  0.100334(6)~&~  0.0045202(7)~
\cr \botrule
\end{tabular}
\label{ltable}
\end{table*}

\subsection{Tabulated numerical data}
\label{sec:resfinite}

All the computed finite-size data for the quantities defined in Sec.~\ref{sec:defs} are listed in Table \ref{ltable}. The length of the SSE runs generating
these results were adapted to achieve a statistical error less than $3 \times 10^{-8}$ in the energy (rounded to one digit in the table). The other quantities
have larger errors. In some cases similar errors were achieved in previous studies, e.g., in Ref.~\cite{sandvik10a} the staggered magnetization was
computed using a valence-bond projector QMC method for sizes up to $L=256$, with errors only somewhat exceeding those in Table \ref{ltable} for the
larger common system sizes. The results of the two calculations agree perfectly within statistical errors. None of the quantities in Table \ref{ltable} have
previously been available at the level of statistical precision achieved here. They will hopefully be useful for benchmarking beyond just the energy.

The results for $L \le 30$ represent averages over simulations at $\beta/L=32$ and $64$. The averaging is validated by the fact that the results agree to within
statistical errors as well as the behaviors of the excitation probabilities graphed in Fig.~\ref{p0_beta}. There is good statistical agreement between most of
the data at the two $\beta$ values also for the larger systems, with the exception of staggered susceptibility $\chi_s$, which will be further discussed below.
The good agreement for all other quantities serves as a validation of the $\beta/L=64$ data being fully converged within the error bars.

In the case of $\chi_s$, results at $\beta/L=32$ and $64$ agree within statistical errors only for $L \le 60$. For larger systems the differences are still minor,
e.g., for the largest system (and deviation) $L=96$ the $\beta/L=64$ value is about $2\%$ larger. Given this small difference, the asymptotically expected
exponential convergence with $\beta$, and the size dependence of the ground state probabilities in Fig.~\ref{p0_beta}, the $\beta/L=64$ values should be fully
converged within statistical errors for all $L<80$. Limited results (with somewhat lower statistical precision) at $\beta/L=128$ for $L=80,88$, and $96$ 
also confirm close convergence. For $L=96$, $\chi_s=0.0045216(13)$, which agrees within the statistical errors with the value in Table \ref{ltable}. The
remaining finite-$T$ deviations in Table \ref{ltable} should therefore be at most a few error bars for $L=96$ and insignificant for the smaller
sizes.

For $L=6$, the ground state energy to nine significant digits is $e_0=0.678872150$ from Lanczos exact diagonalization and the staggered magnetization
squared is $m_s^2=0.209837151$. The corresponding results in Table \ref{ltable} are statistically fully consistent with these values, thus demonstrating the
unbiased nature of the SSE method. For $L=10$, a very precise SSE result was recently obtained, $e_0=-0.67155267(5)$ \cite{chen24}, where the relative error
as small as $10^{-7}$ was needed in order to test an even more precise variational NN results; $e_0=-0.67155260(3)$. These results also agree perfectly with
the $L=10$ energy in Table \ref{ltable}, where the statistical error is even smaller.

Recently an approach based on recurrent NNs was used to compute
energies for $L$ up to $32$, with claimed errors in the seventh decimal place \cite{moss25}. In principle this method should be variational, but the
reported energies extrapolated to zero variance are in some cases below the values in Table \ref{ltable}, including the $L=10$ system for which the earlier
high-precision NN result quoted above agrees with the present SSE result. The discrepancy demonstrates the difficulties associated with variance
extrapolations, though it is still impressive that the results are good to five decimal places.

\section{Finite-size corrections and Infinite-size extrapolations}
\label{sec:infinite}

The form of the finite-size scaling corrections in ordered antiferromagnets have been derived within spin-wave theory \cite{huse88} as
well as more sophisticated symmetry based approached \cite{neuberger89,fisher89}, in the most refined form to high orders in chiral
perturbation theory \cite{hasen93,niedermayer11}. The general form of all the quantities that will be considered here is a polynomial
in the inverse system size, with some of the low-order coefficients given explicitly in terms of infinite-size ground-state parameters.
Because of the high statistical precison of the data, higher-order terms also have to be included in order to obtain good fits when the
smaller system sizes are included.

Some of the predicted low-order coefficients were tested in Ref.~\cite{sandvik97} with data for system sizes only up to $L=16$. Despite the now modest
lattices, agreement was found to within a few percent for the leading corrections in independent polynomial fits. A collective fit to several quantities,
constrained by the predicted forms, gave statistically more precise values of the coefficients. In some cases, in particular the spin stiffness and
velocity, the extrapolated values deviate somewhat (by several error bars) from later determined values, e.g., in Refs.~\cite{jiang11,sen15}.
In hindsight, the system sizes $L\le 16$ in Ref.~\cite{sandvik97} were too small to completely avoid distorting effects of corrections of
higher order than those included. No such bias has been claimed in the case of the energy, however.

With the much reduced statistical errors and larger range of system sizes, the predicted $T=0$ finite-size corrections can now be tested more
precisely. The three predicted forms that will be considered here are for the energy, the sublattice magnetization, and the staggered susceptibility
\cite{neuberger89,hasen93,niedermayer11}:
\begin{subequations}
\begin{eqnarray}
e_0(L) & = & e_0 - \frac{Ac }{L^3} + \frac{1}{4\chi_\perp}\frac{1}{L^4} + \ldots, \label{form_e} \\
m_s^2(L) & = & m^2_s + \frac{Bm^2_s}{\sqrt{\rho_s\chi_\perp}}\frac{1}{L} + \frac{\mu \ln^\gamma(L/\xi)}{L^2} + \ldots,~~ \label{form_m}  \\
\frac{\chi_s(L)}{L^2} & = & m_s^2 \left ( \frac{2\chi_\perp}{3}+\frac{C}{c}\frac{1}{L} +\frac{C}{\rho_2}\frac{1}{L^2} \right ) + \ldots \label{form_x},~~ 
\end{eqnarray}
\label{fitforms}
\end{subequations}  
where the symbols without $(L)$ refer to the infinite-size values and the constants $A$, $B$, and $C$ are
\begin{equation}
A = 1.437745,~~ B = 0.6207465,~~ C = 0.1918021.
\label{abvalues}
\end{equation}
Note that the second-order term in Eq.~(\ref{form_m}) includes a multiplicative log correction with unknown prefactor
$\mu$ and exponent $\gamma$ \cite{neuberger89}. No logs have been claimed in the other quantities. Note also that the leading correction to the
energy is of order $L^{-3}$ and the factor of the $L^{-4}$ correction also has a definite predicted value in terms of only the transverse
susceptibility \cite{hasen93}. The second-order correction to $\chi_s$ was obtained in Ref.~\cite{niedermayer11}. In the original literature on the
finite-size corrections, many of the coefficients in Eqs.~(\ref{fitforms}) were expressed using the spinwave velocity $c$. Here, by using Eq.~(\ref{chydro})
$c$ has been eliminated in some of the terms for simplified expressions.

Though in principle combined fits to several quantities can be employed \cite{wiese94,sandvik97,jiang11}, here independent fits of all the quantities will be
carried out first. A small number of terms (typically two) beyond those included in Eqs.~(\ref{fitforms}) will be added in order to capture remaining finite-size
corrections for the still relatively small system sizes available here. The consistency of those coefficients that are predicted in Eqs.~(\ref{fitforms}) can
then be tested with the extrapolated $L \to \infty$ quantities (i.e., the zeroth-order coefficients). Satisfactory agreement between the fitted and predicted
coefficients then not only supports the theoretical underpinnings of Eqs.~(\ref{fitforms}) but can also be taken as confirmation the soundness of the
extrapolations; specifically the lack of impact of neglected terms of higher order than the corrections included. Having established the applicability of
Eqs.~(\ref{fitforms}), the precision of the extrapolated $e_0$ and $m_s$ will be further improved by statistical imposition of the values of $\chi_\perp$ and
$\rho_s$ in their leading corrections. In all fits below, the raw data without truncating at the level of the one-digit statistical error (in Table \ref{ltable})
were used for full statistical soundness of the procedures.

\subsection{Uniform susceptibility, spin stiffness, and spinwave velocity}
\label{sub:rxc}

In order to eventually test and use Eqs.~(\ref{fitforms}), we begin by extrapolating $\rho_s(L)$ and $\chi_\perp(L)$ in Table \ref{ltable},
using unconstrained polynomials of fourth order. Marginally good fits are then obtained when including system sizes $L\ge 10$ for $\rho_s$ and $L \ge 8$
for $\chi_\perp$. Increasing the minimum size $L_{\rm min}$ included in the fits to $L_{\rm min}=12$ and $L_{\rm min}=10$, respectively, leads to fully
satisfactory goodness-of-fit per degree of freedom; $\chi^2/N_{\rm dof} \approx 1.1$ (for $\rho_s$) and and $0.89$ (for $\chi_\perp$). The $L \to \infty$
extrapolated values (constants in the fits) are then $\rho_s=0.180752(6)$ and $\chi_\perp=0.065690(5)$. The result from constrained fits to $L \le 16$
data in Ref.~\cite{sandvik97} were $\rho_s=0.175(2)$ and $\chi_\perp=0.0625(9)$, both smaller than the present estimates by about three of their error
bars, reflecting some effects of corrections beyond the only quadratic fits used there. A later SSE result obtained from the quantum rotor states of large
systems by analyzing the effects of an external magnetic field on the uniform magnetization gave $\chi_\perp=0.0659(2)$ \cite{syljuasen02}, which is
consistent with the present result but with much larger error. In Ref.~\cite{jiang11} $\rho_s = 0.18081(11)$ was obtained, likewise consistent
with the present result within its ten times larger error.

The spinwave velocity $c$ is also an important ground state parameter and appears in both Eq.~(\ref{form_e}) and Eq.~(\ref{form_x}).
Extrapolating the finite-size results for $c$ in Table \ref{ltable} with $L_{\rm min}=12$ gives $c=1.65881(7)$ ($\chi^2/N_{\rm dof} \approx 1.1$).
Its value according to Eq.~(\ref{chydro}) with the above $\rho_s$ and $\chi_\perp$ values consistent with the above result but with a marginally
smaller error bar; $c=1.65880(6)$, which will be used below. This value is statistically less precise than and inconsistent with
(deviating by about five error bars from) an indirectly computed value $c=1.65847(4)$ \cite{sen15} obtained with SSE simulations at
different aspect ratios $L/\beta$ and locating the special ratio $(L/\beta)^*$ at which the spatial and temporal winding number fluctuations are
equal. At that point the system is effectively space-time symmetric and $c=(L/\beta)^*$. This method had previously been used with less precise data
in Ref.~\cite{jiang11}, with the result $c = 1.6586(3)$. Given the overall trends of $c(L)$ in Table \ref{ltable}, it is unlikely that the
value could be as low as $c = 1.65847(4)$ obtained in Ref.~\cite{sen15}. Determining the cubic ratio $(\beta/L)^*$ is also more involved \cite{sen15}
than the simple fitting of simple SSE estimators used here, thus implying higher risks of underestimating the errors propagating from all steps. The value
and error bar obtained by the simple extrapolations above are therefore likely more reliable

\subsection{Ground state energy}
\label{sub:e0}

We now proceed to test the form Eq.~(\ref{form_e}) of the ground state energy. The approach taken here is again to use a polynomial of high enough order to
obtain a statistically sound fit even when small system sizes are included. Adding terms of order $L^{-5}$ and $L^{-6}$ in Eq.~(\ref{form_e}), i.e.,
with five adjustable parameters in total, a fit to all the data for $L\ge 6$ from Table \ref{ltable} has excellent goodness-of-fit, $\chi^2/N_{\rm dof}=0.94$.
However, the predicted $L^{-3}$ and $L^{-4}$ coefficients in Eq.~(\ref{form_e}) are not perfectly reproduced. Denoting these coefficients by $e_3$ and $e_4$,
respectively, the $\rho_s$ and $\chi_\perp$ values determined above correspond to $e_3 = 2.38493(8)$ and $e_4=3.8058(3)$, while the fit with minimum size
$L_{\rm min}=6$ produces $e_3=2.3785(6)$ and $e_4=3.54(2)$. Increasing the minimum size to $L_{\rm min}=8$ gives $e_3 = 2.383(2)$, which is consistent with the
predicted value, while $e_4=3.64(5)$ is only marginally consistent, deviating by about three error bars. Increasing the minimum size to $L_{\rm min}=10$, both
parameters are fully consistent with the predicted values but the error bars of course are larger as well; $e_3 = 2.382(3)$ and $e_4=3.7(1)$

\begin{figure}[t]
\includegraphics[width=84mm]{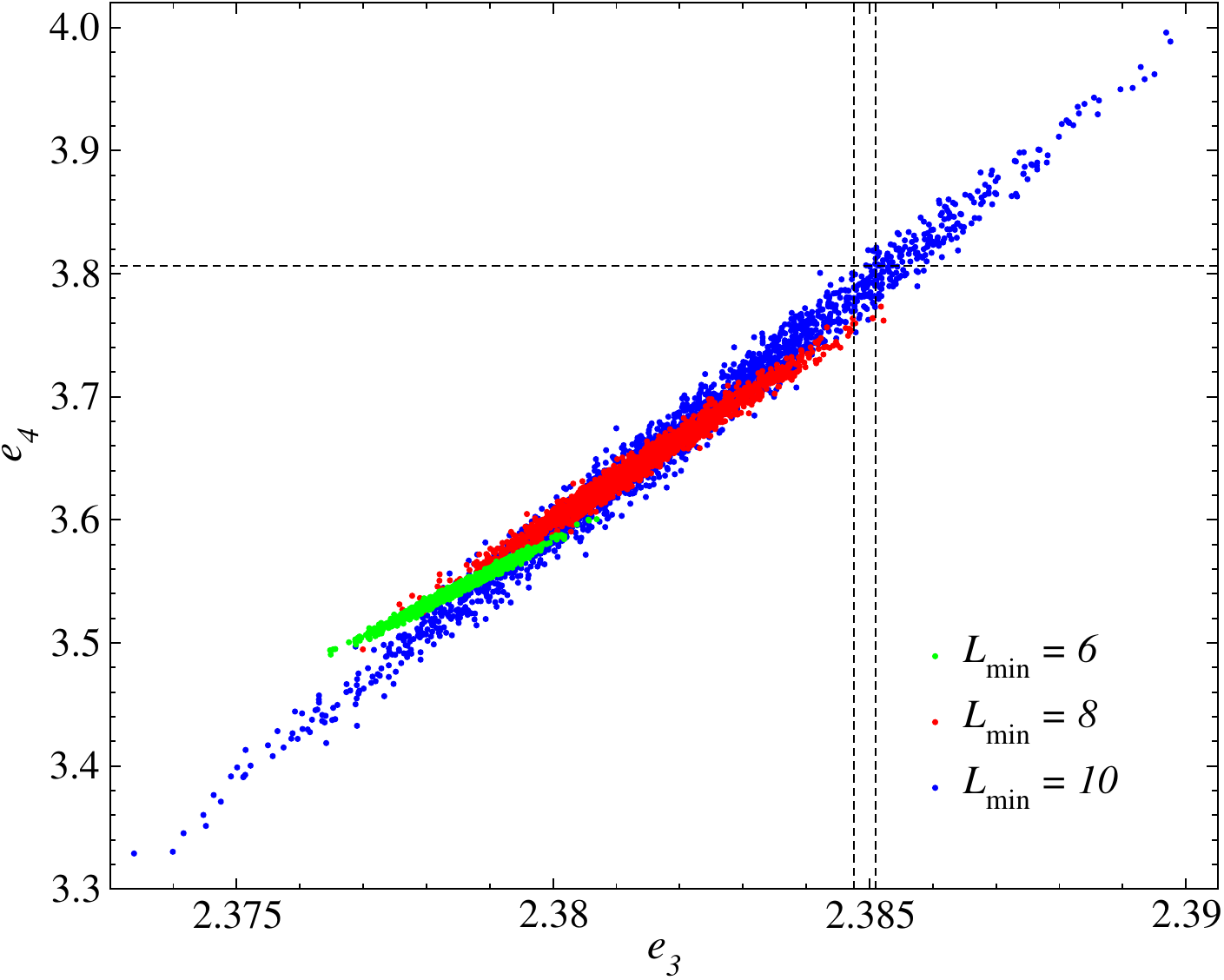}
\caption{Scatter plot of the coefficients of the cubic and quartic corrections of the energy based on 2000 fits with Gaussian noise added to the
energy data in Table.~\ref{ltable}, with the minimum size being $L_{\rm min}=6$ (green dots), $8$ (red dots), and $10$ (blue dots).
The horizontal and vertical pairs of dashed lines indicate the expected values based on Eq.~(\ref{fitforms}) with $\rho_s$ and $\chi_\perp$ taking the
previously extrapolated values, plus and minus two standard deviations in each case. The two horizontal lines almost coincide on the scale
used here.}
\label{scatter1}
\end{figure}

The fitted coefficients are not statistically independent, and for a view of their correlated statistical fluctuations it is useful to examine the full
distribution $P(e_3,e_4)$. Here this distribution is generated by adding Gaussian noise to the energy data in Table \ref{ltable}, with mean zero and
standard deviation equal to the one-standard deviation statistical errors. Results based on 2000 fits with different noise realizations are shown in
Fig.~\ref{scatter1} for $L_{\rm min}=6$, $8$, and $10$. The predicted values of the coefficients are indicated by the vertical and horizontal lines. It is
obvious that the fitted coefficients $(e_3,e_4)$ indeed are highly correlated and that the $L_{\rm min}=6$ fits do not produce valid coefficients. The agreement
is better with $L_{\rm min}=8$, though still not acceptable, with the part of the distribution with significant probability density still slightly outside the
box corresponding to the jointly allowed values.  Only with $L_{\rm min}=10$ do the fits produce a distribution with statistically plausible overlap with the box,
and for $L_{\rm min}=12$ (not shown) the box is well inside the (still larger) high-probability region. The individually computed values with statistical
errors quoted above correspond to projecting the two-dimensional distribution onto the $e_3$ and $e_4$ axes. These distributions are much broader than the thickness
of the very elongated full distributions. Fig.~\ref{scatter1} shows that the agreement with both of the the predicted values for $L_{\rm min}=10$ is actually much
better than indicated by just the independent error bars, in the sense that fixing one of the coefficients to within its predicted window gives a very small error
bar of the second coefficient, and that coefficient is then also consistent with the prediction.

The extrapolated $L\to \infty$ energy for $L_{\rm min}=10$ is $e_0=-0.669441865(11)$ and the results for $L_{\rm min}=8$ and $L_{\rm min}=6$ are also fully
consistent with tis value, $e_0=-0.669441866(10)$ and $e_0=-0.669441878(9)$, respectively. These results are also statistically fully consistent with the often
quoted previously best infinite-size value \cite{sandvik97}, $e_0=-0.669437(5)$, where the error is about $500$ times larger. Given the almost identical
results for $L_{\rm min}=8$ and $L_{\rm min}=10$, the final result of this analysis is taken as $e_0=-0.669441866(10)$.

Given that there should now be no doubt about the applicability of Eq.~(\ref{form_e}) in combination with the higher-order corrections,
an even more precise extrapolated energy can be obtained by constraining the fit by imposing the values of the coefficients $e_3$ and $e_4$ within their
statistical errors. As an alternative to carrying out such a constrained fit, an equivalent method is to use the distribution $P(e_3,e_4)$ and compute
the mean energy from only those fits that produce $e_3$ and $e_4$ values falling within the box marking the allowed statistically acceptable window
(of size one standard deviation in each direction). Sampling millions of points, for $L_{\rm min}=10$ only about $8\times 10^{-3}$\%
of them satisfy the constraints (reflecting the small size of the allowed box inside the much larger region of significant weight in the sampled
distribution) and the mean  goodness-of-fit of the points satisfying the constraint is good, with $\langle \chi^2\rangle /N_{\rm dof}< 1$ (after correcting
for the fact that there are two sources of fluctuations when Gaussian noise is added to data already affected by the statistical errors in the SSE
simulations). The mean energy with $L_{\rm min}=10$ is $e_0=-0.669441861(6)$, while increasing to $L_{\rm min}=12$ changes the result only very little,
to $e_0=-0.669441857(7)$, with $6\times 10^{-3}$\% of the points satisfying the constraints.

Since, it may be argued, the $L_{\rm min}=10$ distribution in Fig.~\ref{scatter1} is not fully satisfactory in relation to the predicted allowed window of
$(e_3,e_4)$, while for $L_{\rm min}=12$ (where the not shown distribution covers a larger area) the allowed window is in a region of high probability density,
the $L_{\rm min}=12$ value  $e_0=-0.669441857(7)$ is here presented as the final result for the ground state energy; the relative statistical error is
unprecedented at $10^{-8}$ and no systematic errors should be expected. Note that the value is only about one error bar smaller than in the plain
extrapolated $L_{\rm min}=10$ and $L_{\rm min}=12$ results; thus there is full statistical consistency but the error bar with the constraint imposed
is about $30$\% smaller.

\subsection{Staggered magnetization}
\label{sub:ms}

When fitting data for the staggered magnetization, the results for larger system sizes from Ref.~\cite{sandvik10a} will
also be included. These are
\begin{eqnarray}
  m^s_s(\hbox{$L=128$}) & = & 0.098843(16),\nonumber \\
  m^s_s(\hbox{$L=192$}) & = & 0.097371(11), \\
  m^s_s(\hbox{$L=256$}) & = &0.096669(17). \nonumber
\end{eqnarray}
In addition to marginally improving the final estimate of $m_s$, these values from a different calculation also serves as an independent consistency test.
The two best previous estimates of the $L\to \infty$ order parameter are both $m_s=0.30743(1)$ \cite{sandvik10a,jiang11}. The result in Ref.~\cite{sandvik10a}
was obtained by extrapolating the long-distance ($r=L/2$) spin correlation function, which has less size dependence than the space-integrated correlator $m_s^2$.
In Ref.~\cite{jiang11}, a collective fit to temperature and size dependent data was used with constraints from chiral perturbation theory.

Before considering the predicted log correction of the quadratic term in Eq.~(\ref{form_m}), it is interesting to explore purely polynomial fits to $m_s^2(L)$.
The log correction was included previously in Ref.~\cite{runge92a} (though with data from a biased QMC method) but not in Refs.~\cite{sandvik97}
and \cite{sandvik10a}. A fourth-order polynomial fits the $L \le 12$ data well, with $\chi^2/N_{\rm dof} \approx 1.1$ (while for $L_{\rm min}=10$ the value is
slightly too high at $\chi^2/N_{\rm dof} \approx 1.9$), and gives $m_s=0.30739(1)$.
This result is only marginally consistent with the previous estimates and has a
similar error bar; with two significant digits in the error the result is $m_s=0.307387(10)$. Reducing the order of the polynomial to third order, a good
fit can only be obtained when the minimum size is increased to $L_{\rm min}=20$ ($\chi^2/N_{\rm dof} \approx 1.0)$, reflecting the sensitivity of the fit to
the high-order corrections that are manifested in the small-size data with very small error bars. The result is then $m_s=0.307394(12)$, fully consistent
with the fourth-order result and  marginally consistent with the previous best results.

Returning to the fourth-order fit with  $L_{\rm min}=12$, even though $\chi^2/N_{\rm dof}$ is good the first-order coefficient $m_1=0.5463(6)$, is
rather far from the predicted value $m_1 = Bm^2_s(\rho_s\chi_\perp)^{-1/2}=0.53826(4)$ when the constant $m^2_s$ from the fit is used along with the previously
determined $\rho_s$ and $\chi_\perp$ values. The agreement is also not improved if the minimum size $L_{\rm min}=14$ is used, which still gives
an excessively large coefficient $m_1=0.5447(9)$ while the sublattice magnetization $m_s=0.30741(13)$ remains consistent with the $L_{\rm min}=12$ value.
Increasing to $L_{\rm min}=16$ gives $m_1=0.5439(12)$, still too large, while the sublattice magnetization increases slightly to $m_s=0.30742(16)$,
which is fully compatible with the results of  Refs.~\cite{sandvik10a,jiang11} but with a larger error. The persistence of a discrepancy
between $m_1$ from the fits and its predicted value can be taken as a indication of the pure polynomial not being the correct function 
to describe the $m_s^2(L)$ data.

\begin{figure}[t]
\includegraphics[width=80mm]{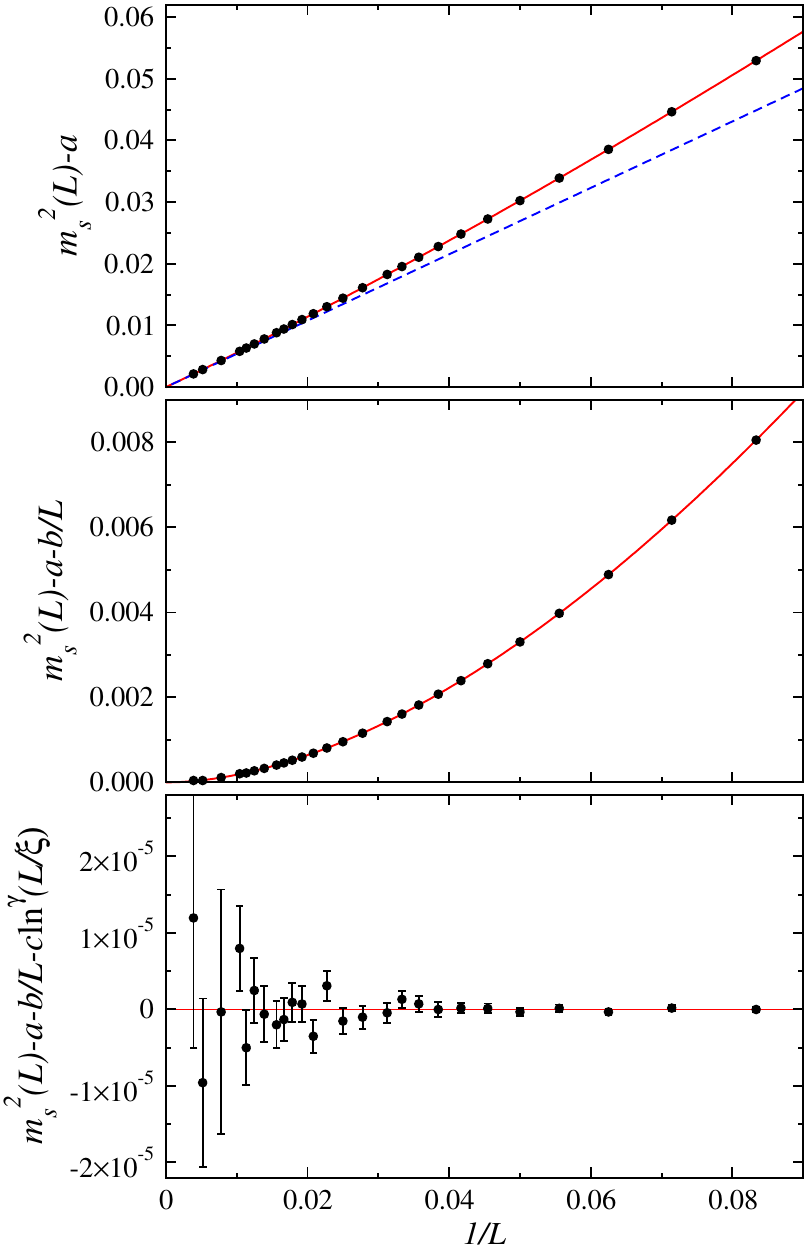}
\caption{The squared sublattice magnetization with the fitted terms of Eq.~(\ref{form_m}) gradually removed from the data set and the
  corresponding remaining functions. No higher-order terms were included. In (a), only the constant $a=m^2_s(\infty)$ has been removed. The
  solid red curve shows the remaining part of the fitted function and the blue dashed line represents the term $b/L$, where the coefficient $b$ is given
  by Eq.~(\ref{form_m}) with the previously infinite-size extrapolated values of $\rho_s$ and $\chi_\perp$. In (b) also the linear term has been removed
  from both the data set and the fitted function. In (c) the remaining $\mu\ln^\gamma(L/\xi)L^{-2}$ term has been removed. There are no sign of higher-order
  corrections within the statistical errors.}
\label{logm}
\end{figure}

Fitting with the predicted log correction included, the values of $\rho_s$ and $\chi_\perp$ in the linear coefficient in Eq.~(\ref{form_m}) are also fixed at their
previously determined values, with their error bars taken into account by repeated fits with Gaussian noise. The constant $m^2_s$ and the linear coefficient are
then constrained. In the log correction of the $L^{-2}$ coefficient,
the factor $\mu$, the exponent $\gamma$, and the scale factor $\xi$ are all unknown and used as free fitting parameters.
Thus, there are four independent parameters and at this stage no higher-order polynomial terms (that might be expected) are included. For $L_{\rm min} = 8$,
this fit is not statistically acceptable, the goodness-of-fit being $\chi^2/N_{\rm dof} \approx 3$. For $L_{\rm min} = 10$, the quality of the fit is already
reasonably good with $\chi^2/N_{\rm dof} \approx 1.3$.

Recalling that the quality of the fourth-order polynomial fit was poor for $L_{\rm min}=10$, the fact that a good fit is now obtained without the $L^{-3}$ and
$L^{-4}$ terms lends strong support to the presence of the multiplicative log correction, and, moreover, suggests that higher-order corrections must be very
small. Excellent fits are obtained for $L_{\rm min}=12$ and higher, with $\chi^2/N_{\rm dof}$ in the range  $0.7 \sim 0.8$ for  $L_{\rm min} \in [12,20]$.
The order parameters obtained from the acceptable fits are remarkably stable for $L_{\rm min}\ge 12$, with $m_s=0.307451(2)$ ($L_{\rm min}=10$),
$0.307444(3)$ ($L_{\rm min}=12$), $0.307443(3)$ ($L_{\rm min}=14$), $0.307443(3)$ ($L_{\rm min}=16$), $0.307443(4)$ ($L_{\rm min}=18$),
and $0.307446(7)$ ($L_{\rm min}=20$). The only slowly increasing error bar here reflects the fact that the log corrected $L^{-2}$ term decays rather
slowly and remains important also for the largest $L_{\rm min}$ sizes considered here. Given the fast convergence, the $L_{\rm min}=12$ result $m_s=0.307444(3)$
is taken as the final result of this analysis. This result agrees perfectly with those of Refs.~\cite{sandvik10a,jiang11}, but with an error only
a third of the previous size.

\begin{figure}[t]
\includegraphics[width=80mm]{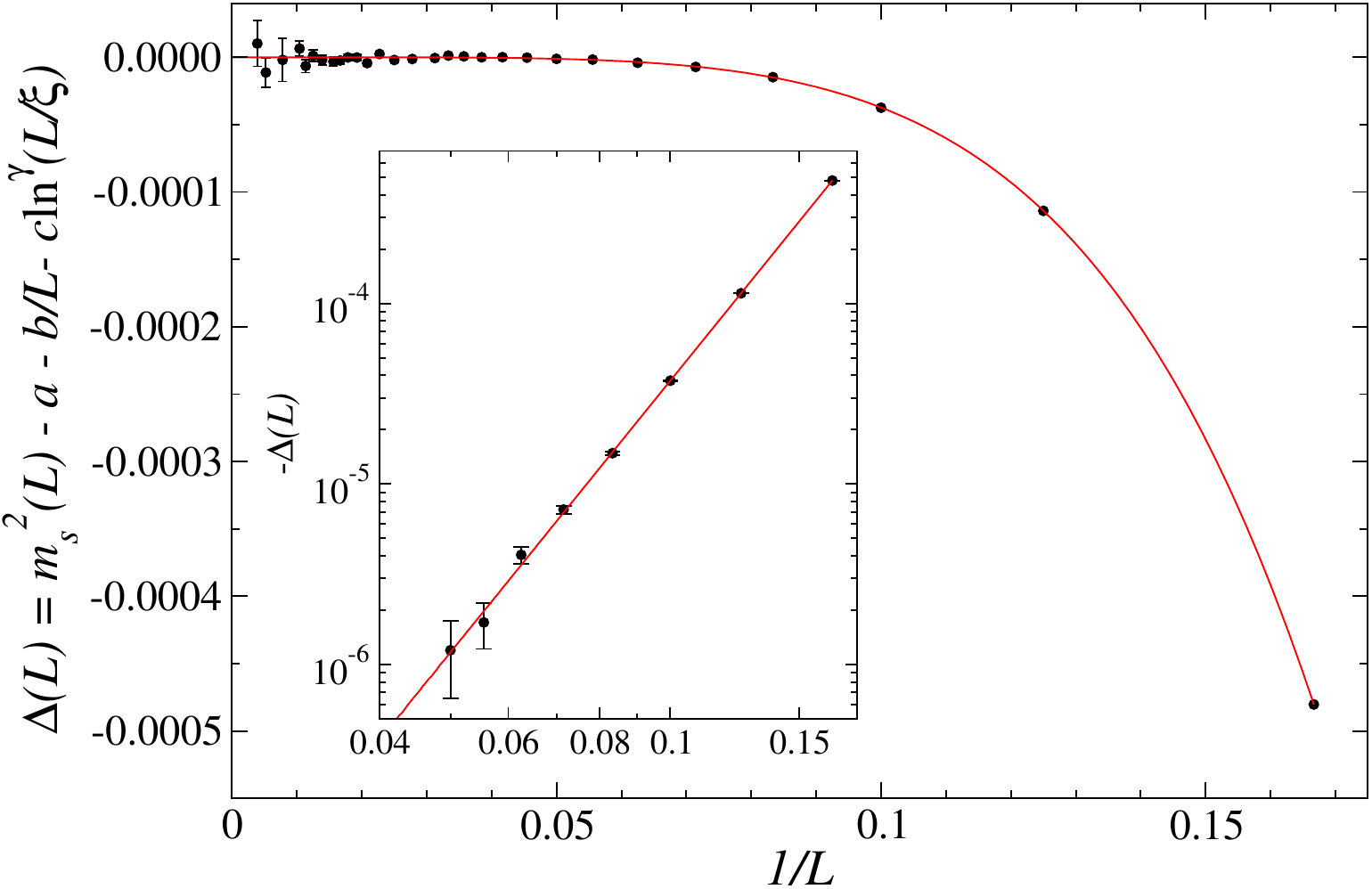}
\caption{Residual $\Delta(L)$ of the squared magnetization when all contributions from Eq.~(\ref{form_m}) have been subtracted (in the corresponding expression
on the $y$-axis, the fitting parameters are denoted $a$, $b$, $\xi$, and $\gamma$). This procedure leaves only the corrections, which in this fit is an $L^{-5}$
term, shown as the red curve, added to Eq.~(\ref{form_m}). The goodness-of-fit is $\chi^2/N_{\rm dof} \approx 0.85$. The inset shows the same residuals (with
a minus sign) in a log-log plot for the points that are clearly different from zero within error bars.}
\label{logm5}
\end{figure}

The value of the exponen $\gamma$ on the log correction in Eq.~(\ref{form_m}) was not predicted in Ref.~\cite{neuberger89}. In the  $L_{\rm min} = 10$
fit, $\gamma=0.73(1)$. For $L_{\rm min}=12$ the exponent increases slightly but then remains very stable within its statistical error: $\gamma=0.82(3)$ for
$L_{\rm min}=12$, $0.83(4)$ for $L_{\rm min}=14$, $0.81(5)$ for $L_{\rm min}=16$, $0.83(8)$ for $L_{\rm min}=28$, and $0.70(14)$ for $L_{\rm min}=20$. Considering
the stability of the $\gamma$ values for $L_{\rm min} \ge 12$, an error weighted average of the values can be taken as the final estimate; $\gamma=0.82(4)$.
The scale factor $\xi$ in the log is also stable with respect to $L_{\rm min}$, with $\xi =0.60(4)$ an overall representative result.

The good fit to $m^2_s(L)$ without higher-order corrections in Eq.~(\ref{form_m}) is further illustrated in Fig.~\ref{logm}. The
raw $L \ge 12$ data with the fitted constant $m^2_s$ subtracted off is shown in Fig.~\ref{logm}(a)  along with the corresponding fitted function (red curve).
The contributions from only the linear term is also shown separately (dashed blue line). In Fig.~\ref{logm}(b) the linear term has also been subtracted, i.e.,
the function shown along with the data is only the quadratic term with its multiplicative log correction. Fig.~\ref{logm}(c) shows the SSE data with the
full fitted function subtracted. The remaining values now fluctuate around zero in a way statistically consistent with Gaussian noise, as would also be
expected because of the small $\chi^2/N_{\rm dof}$ value quoted above. While some higher-order corrections to Eq.~(\ref{form_m}) should of course exist,
these results suggest that at least the $L^{-3}$ correction should be extremely small or even completely absent.

Assuming that the higher-order corrections are pure integer powers of the inverse system size, $a_n/L^n$ with $n \ge 3$, the smallest $n$ for which the
prefactor $a_n$ is significant can be estimated by including the system sizes that were too small in the above fits using only the terms in Eq.~(\ref{form_m}).
When including all system sizes $L \ge 6$, a good fit cannot be obtained for $n=3$. For $n=4$ the fit is better, but still not statistically
sound, with $\chi^2/N_{\rm dof} \approx 4$. However, a very good fit with $\chi^2/N_{\rm dof} \approx 0.85$ is produced when $n=5$. The extrapolated
sublattice magnetization in this case is $m_s=0.307447(2)$, which is consistent with the results without the added correction but with a slightly smaller
error. The exponent on the multiplicative log, $\gamma=0.84(2)$, is likewise consistent with the previous result from the fit without $L^{-5}$ correction,
again with a reduced error. Thus, it appears that the $L^{-5}$ term accounts well for the leading corrections to Eq.~(\ref{form_m}). The other two
parameters of the log term in the fit with this correction are $\xi=0.56(1)$ and $\mu=0.46(1)$.

The SSE data points after subtracting off the terms included in Eq.~(\ref{form_m}) are shown in Fig.~\ref{logm5} along with the corresponding
remaining part of the fitted function, i.e., the pure $L^{-5}$ term. The correction has clearly visible impact for $L \alt 20$, though small changes
in the other fitting parameter can account for the very small effects for $L \ge 12$, as demonstrated by the fact that the residuals are statistically
zero in the fit without correction, Fig.~\ref{logm}(c). While the above analysis of the $L^{-5}$ correction does not necessarily mean that $L^{-3}$
and $L^{-4}$ corrections are completely absent, they at least have to be very small relative to the $L^{-5}$ term---possibly because of some
cancelation effects. It is of course also possible that at least some of the higher-order terms also have log corrections.

\subsection{Staggered susceptibility}

The predicted form of the size-normalized staggered susceptibility $\chi_s$(L) in Eq.~(\ref{form_x}) is qualitatively different from the energy and the staggered
magnetization in that also the constant term has a predicted form, depending on the above determined staggered magnetization in addition to the uniform
susceptibility. Thus, the aim here is not to determine the constant but to test its value against the predicted coefficient computed from the
$L \to \infty$ values of $m^2_s$, $\chi_\perp$, and $c$ determined above. With three coefficients $(x_0,x_1,x_2)$, the distribution $P(x_0,x_1,x_2)$
from repeated fits is of interest. For easier visualization, the distribution will here be projected down to the two-dimensional distributions
$P(x_0,x_1)$ and $P(x_1,x_2)$, based on sampling 2000 fits with Gaussian noise added to the $\chi_s(L)$ SSE data in Table \ref{ltable}.

\begin{figure}[t]
\includegraphics[width=80mm]{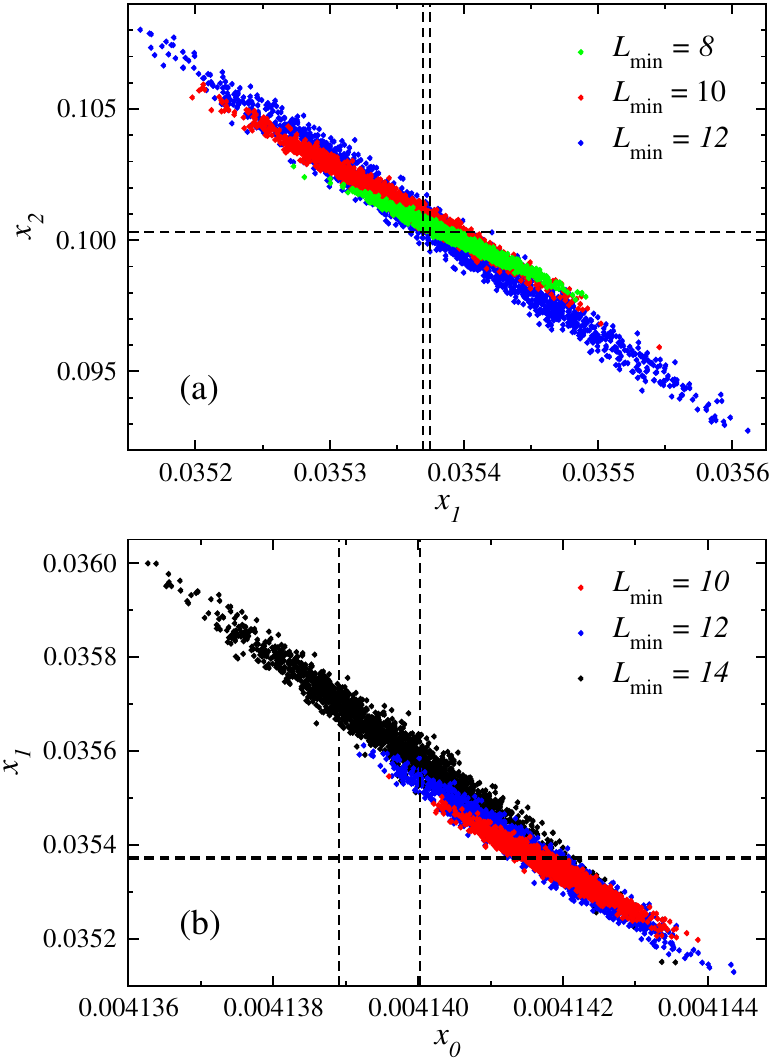}
\caption{(a) Scatter plot of the coefficients of the linear ($x_1$) and quadratic ($x_2$) corrections of the staggered susceptibility, based on 2000 fits
with Gaussian noise added to the $\chi_s$ data in Table.~\ref{ltable}, with the minimum size $L_{\rm min}=8$ (green dots), $10$ (red dots), and $12$ (blue dots).
The horizontal and vertical pairs of dashed lines (the horizontal lines almost coincide) indicate the expected values based on Eq.~(\ref{form_x}) plus and minus
two standard deviations, with $\rho_s$, $c$, and $m_s^2$ set to their previously determined values. (b) Similar plot of the constant term ($x_0$) and the
coefficient of the linear correction ($x_1$) obtained with $L_{\rm min}=10$ (red), $L_{\rm min}=12$ (blue), and $L_{\rm min}=14$ (black).}
\label{scatter2}
\end{figure}

The linear and quadratic coefficients, $x_1$ and $x_2$, are tested first in Fig.~\ref{scatter2}(a) with data points generated using fourth-order polynomial
fits with $L_{\rm min}=8$, $10$, and $12$, all of which have $\chi^2/N_{\rm dof} < 1$. Only marginally good consistency with the predicted window of coefficients is
obtained with $L_{\rm min}=8$ and $10$, but for $L_{\rm min} = 12 $ (and larger) the window is in a region of high probability density $P(x_1,x_2)$. However, the
distribution $P(x_0,x_1)$ is far less convincing, as shown Fig.~\ref{scatter2}(b) for $L_{\rm min}=10$, $12$, and $14$. While the individual $x_0$
and $x_1$ windows  do have large overlaps with the distribution, the joint $(x_0,x_1)$ window is far away from the high-probability region, to the extent that
an agreement is statistically very unlikely (noting also that the indicated window corresponds to two standard deviations of the mean coefficients).
The trend with increasing $L_{\rm min}$ is also not indicative of eventual agreement for for $L_{\rm min} > 14$ (until the size of the distribution
grows too large for meaningful comparisons).

As discussed in Sec.~\ref{sec:resfinite}, there may still be some $T> 0$ contributions left in $\chi_s(L=96)$ at $\beta/L=64$, potentially leading to
underestimation of the value by up to a few of its error bars in Table \ref{ltable}. To exclude $T>0$ effects as the cause of the mismatch in
Fig.~\ref{scatter2}(b), fits with one or several large systems excluded were also carried out. In all cases the changes in the distributions are very minor
and the allowed window falls within the distribution only when the fluctuations become too large to produce meaningful results (when many small and large
sizes have been removed).

The behavior here is similar to the case of $m^2_s$ fitted to a polynomial in $L^{-1}$
without the log correction in Sec.~\ref{sub:ms}, where the linear coefficient $m_1$ was not consistent with the predicted form in Eq.~(\ref{form_m}).
The discrepancy was only resolved by including the log correction to the quadratic term. Given that $m^2_s$ and $\chi_s$ are strongly related to each
other in terms of their definitions (being space and space-time integrals, respectively, of the spin correlation function), one might suspect that
also $\chi_s$ should have a log correction, perhaps in the cubic term. The fitted polynomial may then skew the coefficients of the lower-order term
in a similar way as in the case of $m_1$ in the $m^2_s$ fits.

In principle, another potential reason for the discrepancies could be that some of the extrapolated coefficients in Eq.~(\ref{form_m}) have errors beyond the
quoted statistical errors. The most likely candidate for such errors would be in $x_0=(2/3)m_s^2\chi_\perp$, which would fix the problem in Fig.~\ref{scatter2}(b)
if the window is shifted by $x_0 \to ax_0$ with the factor $a \approx 1.0005$. Such a shift would imply the presence of substantial extrapolation errors
in either $\chi_\perp$ or $m_s^2$, or both. In particular, a shift of only $m_s^2$, which is a factor in all the coefficients $x_0,x_1,x_2$ in
Eq.~(\ref{form_m}) would solve the problem in Fig.~\ref{scatter2}(b) and still maintain consistency with Fig.~\ref{scatter2}(a). Such a large extrapolation
error in $m_s$ is unlikely, however, as the value would have to increase to $m_s \approx 0.30752$, which is very far out of the range determined in
Sec.~\ref{sub:ms} and is also about 10 error bars away from the best previous estimates \cite{sandvik10a,jiang11}. The remaining possibility of
a shift in $\chi_\perp$ is also unlikely, based on many different extrapolations performed that deliver results consistent with that reported in
Sec.~\ref{sub:rxc}. Thus, the presence of a logarithmic correction in Eq.~(\ref{form_x}), whether in one of the terms displayed or in some
higher-order correction, remains the most likely explanation for the discrepancy in Fig.~\ref{scatter2}(b).

Given that the pure polynomial fitting functions may not be strictly correct, an independent (of the previously determined coefficients) value
of $\chi_s/L^2$ from $L \to \infty$ extrapolations including the smallest system sizes should be regarded with some suspicion.
even though the $\chi^2/N_{\rm dof}$ values are good. The effects of the unknown corrections should diminish with increasing $L_{\rm min}$. Given
that the $L_{\rm min}=14$ distribution of $x_0$ from Fig.~\ref{scatter2}(b) overlaps well with the predicted window, the corresponding mean value here
is taken as a reliable result; $\chi_s/L^2 = 0.04140(1)$. The value based on $m_1=(2/3)m^2_s\chi_\perp$ (assuming that this indeed equals $\chi_s$)
is still of higher statistical precision, however, at $0.041384(4)$.

\section{Open and cylindrical boundary conditions}
\label{sec:bc}

Some of the often used lattice many-body methods are not efficient for 2D systems with periodic boundary conditions, e.g., cylindrical boundaries are
typically used in DMRG and related matrix-product state methods  \cite{schollwoeck11,stouden13,yang22} while some of the TN-based methods are only practically
applicable with fully open boundaries \cite{murg09,liu18}. With cylindrical boundaries, it is further advantageous to use non-square lattices of size
$L_x\times L_y$ with $L_y$  larger than the size $L_x$ in the periodic direction In some cases the limit $L_y/L_x \to \infty$ can be taken, in which case
the system strictly is one-dimensional, but the 2D limit can in principle (though often not in practice) be studied for increasing $L_x$ to realize
the 2D limit. In other cases it is better to keep the aspect ratio fixed, e.g., with $L_y/L_x=2$. Here some results for open $L\times L$ systems and
cylindrical $L\times 2L$ systems at $\beta/L=32$ will be presented,
but for a more limited set of quantities and at lower statistical precision than achieved above with 
periodic boundary conditions in the preceding sections.
The ground state energy will now be given per interacting bond $N_b$, while the squared sublattice magnetization and the staggered susceptibility
will be normalized by the system volume $N$ as before. Results are listed in Tables \ref{otable} and \ref{ctable}. For the largest systems, the
staggered susceptibility was not computed as much larger $\beta/L$ values would be needed for convergence. The energy and sublattice magnetization
are $T \to 0$ converged.

\begin{table}[t]
\caption{SSE data for $L\times L$ systems with open boundary conditions. The ground state energy is normalized by the number of interaction bonds
$N_b=2(L^2-L)$.}
\begin{tabular}{|r|l|l|l|}
\toprule    ~$L$~& ~~~~~~$E_0/N_{\rm b}$ & ~~~~~~$m_s^2$ & ~~~~~~$\chi_s/L^2$ \cr
\hline
~ 6~&~0.3621133(1)~&~0.164510(1)~&~0.0098278(2)~
\cr \hline
~ 8~&~0.3537355(1)~&~0.135629(2)~&~0.0070608(2)~
\cr \hline
~10~&~0.3492534(1)~&~0.120522(2)~&~0.0058259(2)~
\cr \hline
~12~&~0.3464731(1)~&~0.111577(2)~&~0.0051646(2)~
\cr \hline
~14~&~0.3445836(1)~&~0.105834(2)~&~0.0047687(3)~
\cr \hline
~16~&~0.3432169(1)~&~0.101926(3)~&~0.0045128(3)~
\cr \hline
~18~&~0.3421827(2)~&~0.099160(4)~&~0.0043392(5)~
\cr \hline
~20~&~0.3413732(2)~&~0.097131(4)~&~0.0042171(5)~
\cr \hline
~22~&~0.3407218(2)~&~0.095619(5)~&~0.0041250(6)~
\cr \hline
~24~&~0.3401870(2)~&~0.094454(5)~&~0.0040592(6)~
\cr \hline
~26~&~0.3397397(2)~&~0.093538(6)~&~0.0040081(6)~
\cr \hline
~28~&~0.3393603(2)~&~0.092827(4)~&~0.0039687(8)~
\cr \hline
~32~&~0.3387504(2)~&~0.091825(7)~&~ ---
\cr \hline
~48~&~0.3373629(2)~&~0.09022(2)~&~ ---
\cr \hline
~64~&~0.3366865(2)~&~0.08998(2)~&~ ---
\cr \hline
~96~&~0.33602096(8)~&~0.09031(3)~&~ ---
\cr \hline
128~&~0.33569204(9)~&~0.09076(5)~&~ ---
\cr \botrule
\end{tabular}
\label{otable}
\end{table}

\begin{table}[t]
\caption{SSE data for $L\times 2L$ lattices with cylindrical boundary conditions (periodic in the shorter direction and open
in the longer direction). The ground state energy is normalized by the number of interaction bonds $N_b=2L^2-L$.}
\begin{tabular}{|r|l|l|l|}
\toprule    ~$L$~& ~~~~~~$E_0/N_{\rm b}$ & ~~~~~~$m_s^2$ & ~~~~~~$\chi_s/L^2$ \cr
\hline
~ 4~&~0.3532363(2)~&~0.170293(2)~&~0.0083710(2)~
\cr \hline
~ 6~&~0.3427879(2)~&~0.140311(3)~&~0.0066647(3)~
\cr \hline
~ 8~&~0.3397983(2)~&~0.125772(3)~&~0.0058507(3)~
\cr \hline
~10~&~0.3384370(2)~&~0.117465(3)~&~0.0053900(4)~
\cr \hline
~12~&~0.3376621(2)~&~0.112218(4)~&~0.0051033(6)~
\cr \hline
~14~&~0.3371612(2)~&~0.108676(5)~&~0.0049103(6)~
\cr \hline
~16~&~0.3368096(2)~&~0.106134(7)~&~0.0047720(9)~
\cr \hline
~18~&~0.3365491(2)~&~0.104248(6)~&~0.0046701(8)~
\cr \hline
~20~&~0.3363476(2)~&~0.102801(8)~&~0.0045925(9)~
\cr \hline
~24~&~0.3360555(3)~&~0.10073(2)~&~  ---
\cr \hline
~32~&~0.3357061(3)~&~0.09843(3)~&~  ---
\cr \hline
~48~&~0.3353698(3)~&~0.09638(5)~&~  ---
\cr \hline
~64~&~0.3352053(2)~&~0.09558(6)~&~  ---
\cr \botrule
\end{tabular}
\label{ctable}
\end{table}

\begin{figure}[t]
\includegraphics[width=80mm]{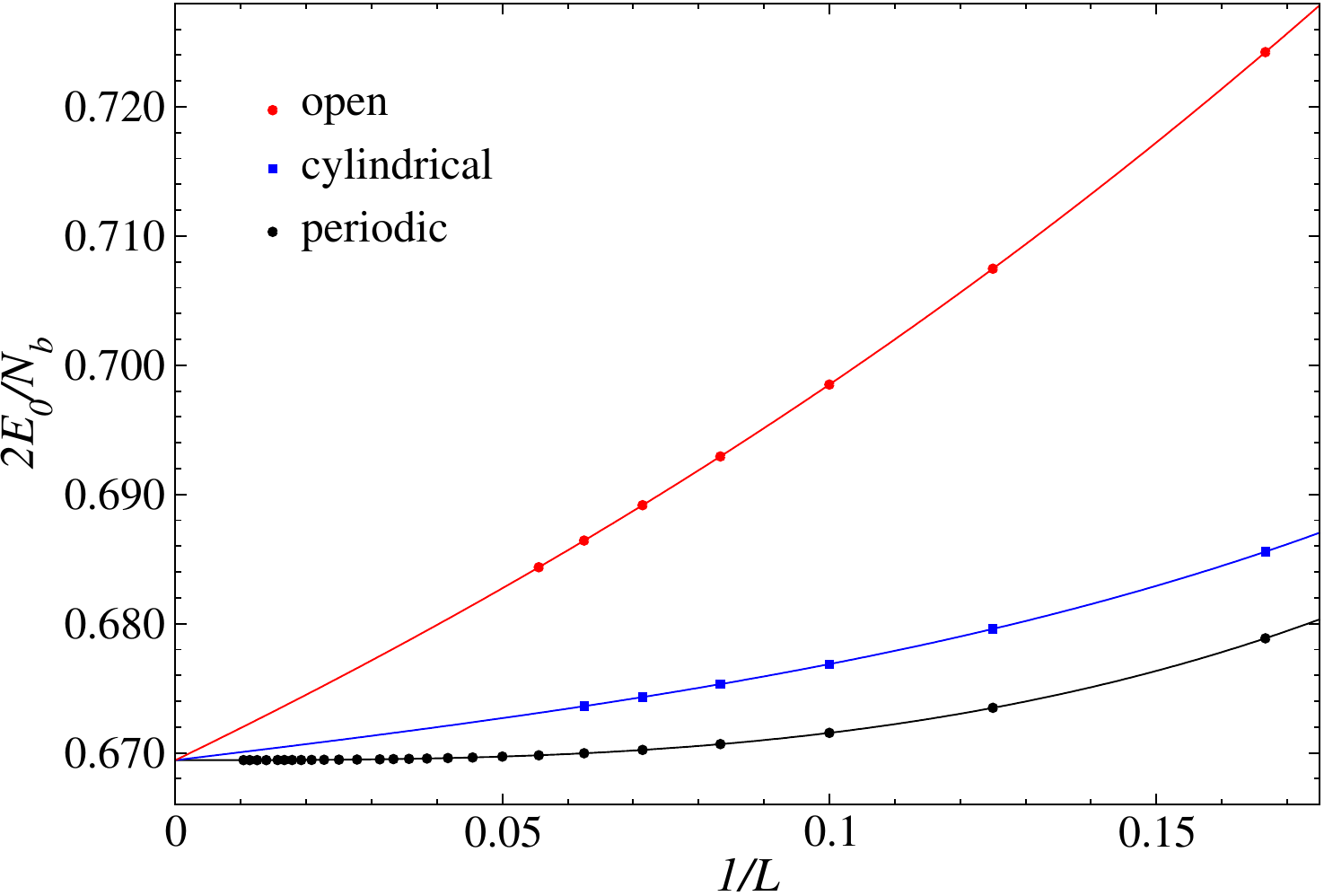}
\caption{Twice the ground state energy per bond for systems with different boundary conditions; periodic $L\times L$ (black circles), cylindrical
$L \times 2L$ (blue squares), and open $L\times L$ (red circles). The curves are polynomial fits, without $L^{-1}$ and $L^{-2}$ terms for the periodic systems but
including those for the other cases.}
\label{e0bc}
\end{figure}

The lack of $L^{-1}$ and $L^{-2}$ terms in the dependence on the energy on the system size in Eq.~(\ref{form_m}) should only be expected with periodic conditions.
With one or both boundaries open, an $L^{-1}$ term should be expected, and there is also no reason for the $L^{-2}$ term to be absent. Indeed, the $E_0/N_b$
data can only be fitted if both these terms are included. Fits are shown in Fig.~\ref{e0bc} for all boundary conditions. For the periodic case, the fit
is the one discussed in Sec.~\ref{sub:e0} and its extrapolated energy was used to constrain the fits for the other two boundary conditions. There are
no predictions for the finite-size corrections for non-periodic systems and the purpose of the fits here is only to demonstrate the $L^{-1}$ and $L^{-2}$
corrections and the expected consistency of the infinite-size energies computed with different boundary conditions.

With fully open boundaries, the convergence of the sublattice magnetization to the previously determined value is not obvious from the
results in Table \ref{otable}. With increasing $L$, $m_s^2(L)$ drops below the expected $L\to\infty$ value and the results for the two largest system sizes
showing the onset of a nonmonotonic behavior before the necessary ultimate convergence to the same value $m^2_s \approx 0.09445$ as for the periodic systems.
The staggered susceptibility as well drops significantly below the results for the periodic systems and nonmonotonic behavior will likely
set in here as well, for system sizes larger than those available. The results for cylindrical systems in Table \ref{ctable} do not indicate
any nonmonotonicity and the results can be extrapolated to values consistent with the periodic lattices.

Suppression of the staggered magnetization close to open boundaries should be expected and the results for $m_s^2$ in Tables \ref{otable} and \ref{ctable}
suggest that the effects are much more dramatic in the case of fully open boundaries. Large fluctuations close to the corners can likely explain the differences.
To more quantitatively investigate the boundary effects on the order parameter, the squared sublattice magnetization can be broken up into site dependent contributions.
Using the normalized sublattice magnetization $m_s^z$ in Eq.~(\ref{mzsdef}), a location dependent squared order parameter can be defined;
\begin{equation}
m^2_s({\bf r})=3\langle S^z({\bf r})m_s^z\rangle(-1)^{r_x+r_y},  
\end{equation}
so that the full (averaged) value is $m_s^2=\sum_{\bf r}m^2_s({\bf r})/N$. A relative distortion field can further be defined as
\begin{equation}\label{distort1}
D(d_x,d_y)=1-\frac{m^2_s({\bf r}_{\bf d})}{m^2_s({\bf r}_{\rm center})},
\end{equation}
where ${\bf d}=(d_x,d_y)$ is the distance to the site at ${\bf r}_{\bf d}$ relative to a reference point on an edge. The normalization is by the squared order
parameter at a site ${\bf r}_{\rm center}$ closest to the center of the system (in practice averaged over all such equivalent sites), where the suppression
(distortion) of the order parameter is the smallest. By definition $d({\bf r}_{\rm center})=0$ at the central sites. $D(d_x,d_y)$ should strictly correspond
to the true distortion field only as $L \to \infty$, where the order parameter at ${\bf r}_{\rm center}$ will truly take its bulk value without finite-size
effects. The vanishing of $D({\bf r}_{\rm center})$ for any $L$ is just a finite-size aspect of the system along with the overall size dependence
of the sublattice magnetization, which is partially compensated for here by using ratios of local order parameters in Eq.~(\ref{distort1}). For the open
$L \times L$ lattices with even $L$ there are four equivalent central locations ${\bf r}_{\rm center}$ while in the cylindrical $L\times 2L$ systems there
are $2L$ such sites (two central ``rings'' of length $L$).

\begin{figure}[t]
\includegraphics[width=75mm]{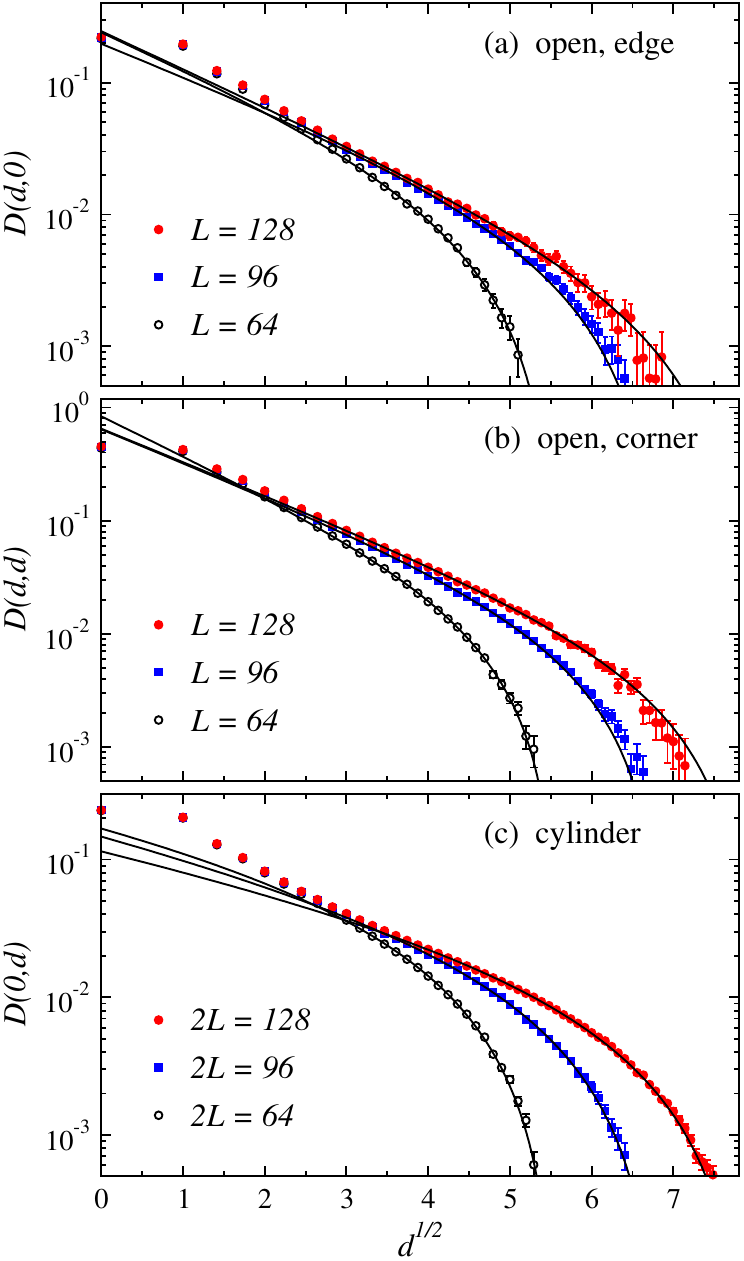}
\caption{Distortion field of the order parameter, Eq.~(\ref{distort1}), vs the square-root of the distance from an edge. In (a), $d$ is the distance
along a line $(d,0)$ extending from a site closest to the mid point of the edge of an open $L\times L$ lattice toward the center of the system, while in (b) the
line extends from a corner in the diagonal direction $(d,d)$. In (c), the distance is from a site on the open edge of length $L$ of an $L \times 2L$ cylinder
along the line $(0,d)$ perpendicular to the open edge. The curves show fits to the stretched exponential decay, Eq.~(\ref{distort2}).}
\label{d_edge}
\end{figure}

In principle it is possible to compute the distortion using linear spinwave theory (which has been applied to a related open-boundary problem
\cite{sanyal12}) or some more sophisticated approach, but this is beyond the scope of the present work. By testing different functional forms, it was
found that a stretched exponential describes the data very well. With $d$ denoting the distance from an open edge or corner  along a line on the lattice
(to be defined more precisely below), the distortion as defined in Eq.~(\ref{distort1}) can be well described by the form
\begin{eqnarray}\label{distort2}
  &&D(d) \propto \left [{\rm exp}\bigl (-\sqrt{d/\xi}\right )+{\rm exp}\left (-\sqrt{(L-1-d)/\xi}\right )\nonumber \\
    &&~~~ -{\rm exp}\left (-\sqrt{(L/2-1)/\xi}\right ) -{\rm exp}\left ( \sqrt{L/2\xi}\right )
    \Bigr ],
\end{eqnarray}
where $L=L_y$ for the fully open lattices and $L=L_y/2$ for the cylindrical systems. The first term is the distortion from the
edge (or corner) with respect to which the distance $d$ is measured and the second term reflects the opposite edge (or corner) at distance
$L-1-r$ (with the distance between the two edges being $L-1$ and the $\sqrt{2}$ factor in a corner-corner diagonal direction is absorbed
into $\xi$). The last two terms ensure the imposed condition $d(r)=0$ in Eq.~(\ref{distort1}) for the center sites located
at $d=L/2-1$ and $d=L/2$. For large $L$ and distances $d \ll L$, only the first term contributes significantly but the other terms are important for
the rather small system sizes and distances considered here.

Results are shown in Fig.~\ref{d_edge} for all three boundary conditions. In the case of the fully open $L\times L$ lattices (with $L$ an even number), the
distance in Fig.~\ref{d_edge}(a) is defined from one of the two sites closest to the mid-point of an edge, extending toward the center of the system along the
line perpendicular to the edge in question. In Fig.~\ref{d_edge}(b), the distance is measured from one of the corners of the open lattice along the
corresponding diagonal direction, not taking into account the factor $\sqrt{2}$ from the length of the diagonal of a plaquette of unit length. 
Fig.~\ref{d_edge}(c) is for the cylindrical system of size $L\times 2L$, with $d$ being the distance from a site on the edge along a line in the
longer direction. In all cases Eq.~(\ref{distort2}) describes the behavior well for sufficiently large $d$. The short-distance deviations
from the form are, perhaps surprisingly, much larger for the cylindrical lattices. In all cases there are still finite size effects in the decay
constant $\xi$, but the trend for increasing system size indicates convergence, with the value for the largest system already close to converged.
For the largest systems, $\xi \approx 1.2$ in Fig.~\ref{d_edge}(a), $\xi \approx 1.3$ in Fig.~\ref{d_edge}(b), and $\xi \approx 2.5$ in Fig.~\ref{d_edge}(c).
The largest distortion, exactly on the edge, is $D(0) \approx 0.22$ in both Fig.~\ref{d_edge}(a) and Fig.~\ref{d_edge}(c), while in the
case of the corner in Fig.~\ref{d_edge}(b) it is about twice as large; $D(0)\approx 0.45$.

\begin{figure}[t]
\includegraphics[width=80mm]{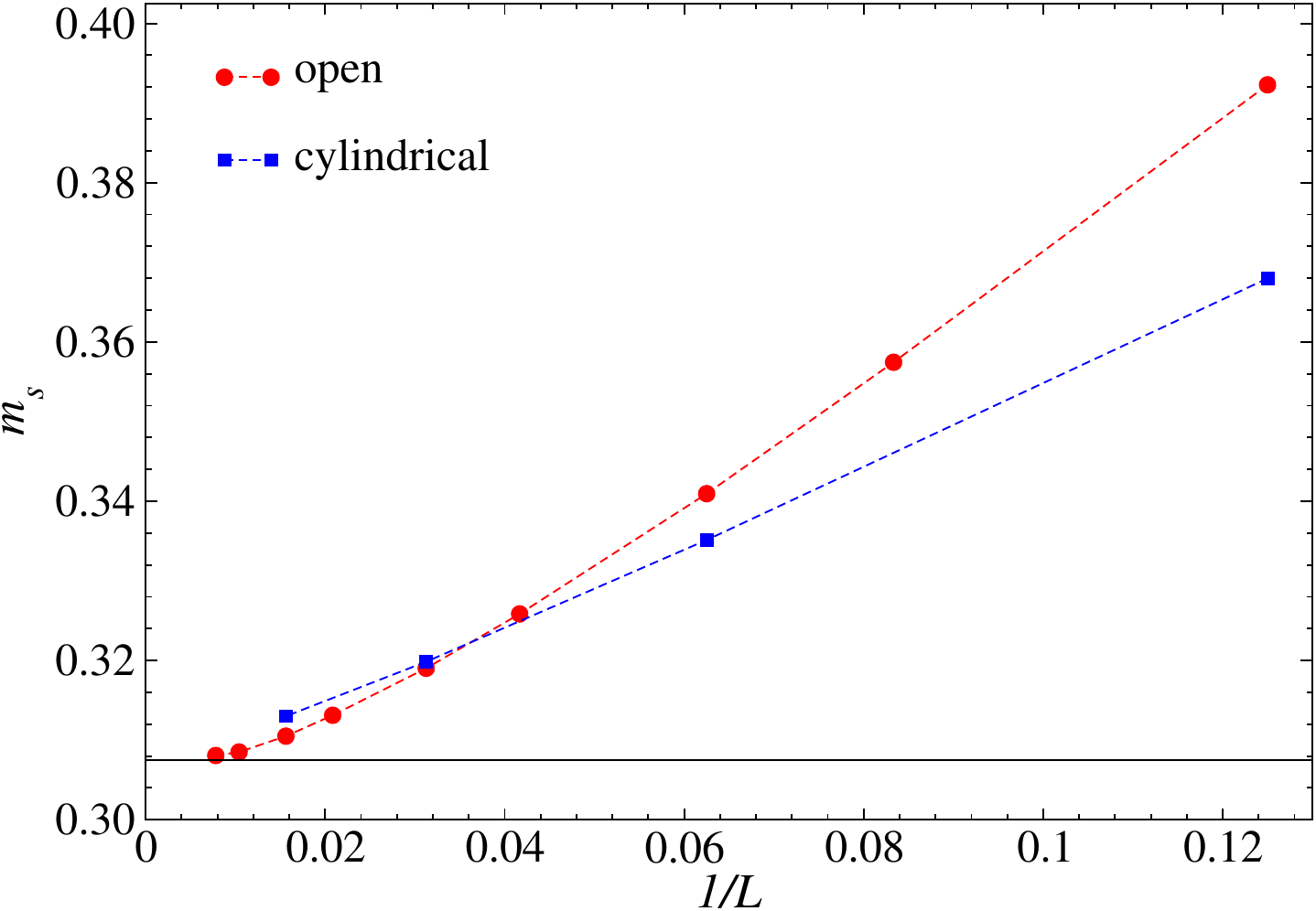}
\caption{Inverse-size dependence of the sublattice magnetization at the sites closest to the center of systems with open (red circles)
and cylindrical (blue squares) boundary conditions. The horizontal line shows the $L \to \infty$ value obtained in Sec.~\ref{sub:ms} for
systems with periodic boundaries.}
\label{m_cent}
\end{figure}

Even though the (stretched) exponential decay guarantees that the spatially averaged $m_s^2$ when $L \to \infty$ will be the same for all three boundary
conditions, the distortion is still relatively large for the moderate lattice sizes considered here, especially because of the corners in the case of the fully
open boundaries. The nonmonotonic behavior seen for the largest system sizes in Table \ref{otable} is then explained as arising from the large
boundary reduction of the order parameter, which affects the spatially averaged value up to large system sizes. The boundary effect also asymptotically
implies a finite-size correction of the form $L^{-1}$ in the average, which can be larger than the linear coefficients in the periodic systems in
Eq.~(\ref{form_m}). This appears to be the case for the open boundaries and the asymptotic approach to the infinite-size value is then from below,
in contrast to the approach from above in the case of the periodic systems.

It is also useful to examine the local staggered magnetization $m_s({\bf r})$ with ${\bf r}$ at the center of the system (averaged over all the
symmetrically equivalent sites closest to the center). Results are shown for both open and cylindrical systems
in Fig.~\ref{m_cent}. Here the results for the cylindrical systems are fully consistent with an approach from above to the expected infinite-size
value. Extrapolations are not shown here as the exact $L \to \infty$ value is sensitive to the functional form used and it is not clear whether
the logarithmic contribution in Eq.~(\ref{form_m}) survives for the cylindrical systems. For the open systems, the results in Fig.~\ref{m_cent}
flatten out for the largest system sizes and a naive extrapolation with a polynomial gives an extrapolated order parameter below the expected
value. Most likely, the behavior here is again nonmonotonic, arising from the open boundaries significantly affecting also the spins in the
center of the system.

\section{Discussion}
\label{sec:discuss}

The N\'eel ordered ground state of 2D Heisenberg $S=1/2$ antiferromagnet represents a corner stone in the field of quantum magnetism. Beyond
specific applications to quantum magnetism, the mechanism of SO(3) spin-rotational symmetry breaking is universal and deserving of detailed
quantitative understanding. A primary motivation for this work was to obtain unbiased high-precision results for the key ground state
parameters of the model, which beyond very small system sizes is possible only with statistically exact QMC algorithms, such as the SSE method
used here. A previously often cited list of ground state parameters up to system size $L=16$ \cite{sandvik97} is outdated,
considering that some of the emergent many-body methods are now capable of exceeding the statistical.
precision reported there, at least in the case of the ground state energy
for $L$ up to $10$ \cite{chen24}. The trend of increasingly good variational states based on TNs and NNs will only continue and better benchmarks
are called for. The raw data listed here in Tables \ref{ltable} (for periodic lattices), \ref{otable} (open lattices), and \ref{ctable}
(cylindrical lattices) are of much higher precision than previously and also cover a broader range of physical observables---including quantities,
such as the staggered susceptibility, that are more challenging to compute with TN and NN approaches than the energy and the order parameter.

The results for the $L \to \infty$ extrapolated ground state parameters are also likely the most precise to date. For ease of reference, the best
estimates of all the quantities considered are repeated here:
\begin{subequations}
\begin{eqnarray}
&&e = -0.669441857(7), \label{val_e} \\
&&\rho_s = 0.180752(6), \label{val_rhos} \\
&&\chi_\perp = 0.065690(5), \label{val_chip}\\
&&c = 1.65880(6), \label{val_c}\\
&&m_s = 0.307447(2), \label{val_m2} \\
&&\chi_s/L^2 = 0.004140(1) \label{val_chis} .
\end{eqnarray}
\end{subequations}
The largest improvement with respect to previous works is in the energy, where the error bar is almost three orders of magnitude smaller than in
Ref.~\cite{sandvik97}. This improvement in precision is not surprising, considering the growth of computing power over almost 30 years in combination
with the improvements in the SSE algorithm, in particular the introduction of loop updates \cite{evertz93,sandvik99}. The algorithmic improvements are
more important for some of the other quantities, however, since the SSE energy estimator is rotationally invariant and not sensitive to the slowest
mode of the stochastic evolution of the SSE configurations (even with inefficient local updates \cite{sandvik97}), which physically correspond to
long-wavelength fluctuations of the order parameter and its overall rotation in spin space \cite{sandvik97}.

In recent work with recurrent NNs, ground state energies for $L \le 32$ were extrapolated to $e_0=-0.6694886(5)$ in the thermodynamic limit \cite{moss25}.
This value differs from the present result above by almost 100 of its error bars, likely primarily because of the non-variational errors introduced by
extrapolating the finite-size results to zero sampling variance. Though the result is only good to the fourth decimal place, the current activities in
pushing NNs to unprecedented precision neverthelss illustrates the increasing needs for reliable benchmarks, such as the extrapolated results above
and the raw finite-size data listed in tables in the preceding sections.

A second purpose of this work, beyond determining the ground state parameters to unprecedented precision, was to test finite-size corrections
predicted by chiral perturbation theory \cite{neuberger89,hasen93,niedermayer11}, which with the improved data was accomplished at a level far
exceeding previous works. A multifaceted statistical analysis showed complete agreement with leading and subleading corrections to both the energy
and the sublattice magnetization. In the latter case, the first subleading ($L^{-2}$) correction is modified by a multiplicative log \cite{niedermayer11},
$\ln^\gamma(L)$, the presence of which was confirmed. The previously unknown value of the exponent $\gamma = 0.82(4)$ was determined.
The corrections to the related staggered susceptibility also show evidence of a log correction in higher-order terms, as evidenced from minor
discrepancies between coefficients from polynomial fits and their predicted forms in terms of other ground state parameters. This result should serve
as a motivation to extend the chiral perturbation theory to higher order---clearly not an easy task, considering the already very sophisticaled
calculations underlying the existing predictions in Eqs.~(\ref{fitforms}).

With open and cylindrical boundary conditions, the reduction of the order parameter at the open edges was found to follow
a stretched exponential form asymptotically. For large sufficiently systems, the order parameter at distance $r$ from an edge takes the form
$m_s(r) = m_s(\infty) - A{\rm exp}(-\sqrt{r/\xi})$, with the parameters $A$ and $\xi$ depending on the reference point on the boundary (specifically, its
distance from corners). In the Heisenberg model the maximum distortion $A$ is rather large, up to almost half of $m_s(\infty)$, and $\xi$ of order one.
The overall edge effects therefore significantly reduce the order parameter integrated over the entire system, in some cases even leading to nonmonotonic
finite-size behavior. These effects should be kept in mind when working with non-periodic boundaries, as data for a range of small system sizes can give
misleading results---this will be the case also if only a central region of the system is used, if the size of that region grows with $L$. Even if only the most
central spins are used, the edge effects can still be important; likely even more so in systems with weaker N\'eel order than the standard Heisenberg
model studied here.

As an example of the computational effort involved to generate the data presented here, in the case of the $L=96$ system at inverse temperature $\beta=64L$,
one full SSE updating sweep with measurements of observables required about $10$s using Intel Xeon Gold 6000-series CPUs. Approximately $1.5 \times 10^{8}$
such steps were performed, for a total of about $4 \times 10^{5}$ hours of CPU core time, distributed over a large number of CPU cores. In the case
of $L=8$ at $\beta=32L$, an updating sweep takes about $10^{-3}$ s and about $2 \times 10^{11}$ steps were carried out for a total of about
$5\times 10^4$ core hours (and this accounted for about half of the statistics for this system size, the remainder coming from $\beta=64L$ runs).

\begin{acknowledgments}
The author would like to thank Uwe-Jens Wiese for discussions. This work was supported by the Simons Foundation under Grant No.~511064. The numerical
calculations were carried out on the Shared Computing Cluster managed by Boston University’s Research Computing Services.
\end{acknowledgments}

\end{document}